%% file: main.tex
\documentclass[sigconf, screen]{acmart}
\acmConference[ISSTA 2024]{ACM SIGSOFT International Symposium on Software Testing and Analysis}{16-20 September, 2024}{Vienna, Austria}
\AtBeginDocument{%
  \providecommand\BibTeX{{%
    \normalfont B\kern-0.5em{\scshape i\kern-0.25em b}\kern-0.8em\TeX}}}

\usepackage{tikz}
\usepackage{amsmath}
\usepackage{listings}

\input{config.tex}

\usepackage{hyperref}
\hypersetup{colorlinks,allcolors=black}

\begin{document}

%don't want date printed
\date{}

\title{\sysname: Learning Transferable Binary Code Representations with Natural Language Supervision}

% Contrastive Language-Assembly Pre-training to boost Transferable Assembly Code Representations

% \title{\sysname: Contrastive Language-Assembly Pre-training for Assembly Code Representation Learning}

%for single author (just remove % characters)
% \author{
% {Anonymous authors}
% % {\rm Second Name}\\
% % Second Institution
% % copy the following lines to add more authors
% % \and
% % {\rm Name}\\
% %Name Institution
% } % end author

\author{Hao Wang$^1$, Zeyu Gao$^1$, Chao Zhang$^1$, Zihan Sha$^2$, Mingyang Sun$^3$, Yuchen Zhou$^4$, Wenyu Zhu$^1$, Wenju Sun$^5$, Han Qiu$^1$, Xi Xiao$^5$}
\renewcommand{\authors}{Hao Wang, Zeyu Gao, Chao Zhang, Zihan Sha, Mingyang Sun, Yuchen Zhou, Wenyu Zhu, Wenju Sun, Han Qiu, Xi Xiao}
\affiliation{%
  \institution{$^1$Tsinghua University, Beijing, China \quad
  $^2$Information Engineering University, Zhengzhou, China \\
  $^3$University of Electronic Science and Technology of China, Chengdu, China  \\
  $^4$Beijing University of Technology, Beijing, China \quad
  $^5$Tsinghua University, Shenzhen, China}
  \country{}
} 
\email{{hao-wang20,gaozy22,zhuwy19,swj22}@mails.tsinghua.edu.cn,{chaoz,qiuhan}@tsinghua.edu.cn}

\email{technicalgrit@foxmail.com,2020090918021@std.uestc.edu.cn,zhouyuchen@emails.bjut.edu.cn}

\email{xiaox@sz.tsinghua.edu.cn}

\renewcommand{\shortauthors}{Wang, et al.}



\begin{abstract}
% In binary analysis, existing deep learning methods for assembly 
Binary code representation learning has shown significant performance in binary analysis tasks. But existing solutions often have poor transferability, particularly in few-shot and zero-shot scenarios where few or no training samples are available for the tasks. 
% These scenarios demand models to efficiently adapt to new tasks with limited or no prior training examples, posing a significant challenge.
% Few-shot learning involves adapting models to new tasks with limited data, while zero-shot learning requires models to handle entirely unseen tasks. 
To address this problem, we present \sysname (\underline{\texttt{C}}ontrastive \underline{\texttt{L}}anguage-\underline{\texttt{A}}ssembly \underline{\texttt{P}}re-training), which employs natural language supervision to learn better representations of binary code (i.e., assembly code) and get better transferability.
At the core, our approach boosts superior transfer learning capabilities by effectively aligning binary code with their semantics explanations (in natural language), resulting a model able to generate better embeddings for binary code.
To enable this alignment training, we then propose an efficient \textit{dataset engine} that could automatically generate a large and diverse dataset comprising of binary code and corresponding natural language explanations. 
% This dataset enables our assembly encoder to learn rich semantic of assembly code.
We have generated \textbf{195 million} pairs of binary code and explanations and trained a prototype of \sysname. 
The evaluations of \sysname across various downstream tasks in binary analysis all demonstrate exceptional performance. Notably, without any task-specific training, \sysname is often competitive with a fully supervised baseline, showing excellent transferability. 
% For instance, it outperforms jTrans on BinaryCorp without using any of the training data. 
We release our pre-trained model and code at \url{https://github.com/Hustcw/CLAP}. 
% \elsa{todo: Update title inside anonymous github}
% This underscores the potential of \sysname in addressing the challenges of task adaptability and data scarcity in binary analysis, setting a new standard in the field.
% \chao{1. importance of transfer learning? 2. novelty of the solution? current version says too much on dataset. 3. why could text supervision boost transferability?}
\end{abstract}

\maketitle

\section{Introduction} \label{sec:intro}

% Deep Learning in Binary Analysis
Deep learning is effective at numerous binary analysis tasks, including function boundary detection~\cite{shin2015boundary}, binary code search~\cite{yang2021codee}, binary code similarity detection~\cite{wangJTransJumpawareTransformer2022,liu2018alphadiff}, function type inference~\cite{chua2017type}, malware classification~\cite{raff2017malware}, reverse engineering~\cite{lacomis2019dire}, and value set analysis~\cite{guo2019deepvsa}. The success of deep learning in the binary analysis field can be attributed to its powerful representation learning capabilities, which have been proven effective in capturing complex patterns and relationships within the data as well as learning meaningful representations of assembly code.
% Despite the progress made by deep learning techniques in binary analysis, a significant challenge arises in transfer learning, particularly when adapting these models to new tasks with limited data.

Despite the progress made by deep learning techniques in binary analysis, existing methods suffer from certain limitations. 
First, current methods typically require substantial data for retraining when applied to new datasets or tasks, leading to poor performance in scenarios with sparse training samples. This is particularly problematic in few-shot learning scenarios, where a model must adapt to new tasks with minimal examples, and zero-shot learning scenarios, where the model encounters tasks it has never seen during training. 
Second, existing schemes for encoding assembly code often result in losing critical information, 
such as the parameters to call instructions, strings, and variable names. In particular, 
most approaches~\cite{wangJTransJumpawareTransformer2022,liPalmTreeLearningAssembly2021,peiTrexLearningExecution2021,dingAsm2VecBoostingStatic2019} tend to normalize character text, constant numerical values, and external functions using special vocabulary tokens, which inadvertently leads to the omission of vital details.

\iffalse
Existing methods suffer from certain limitations. First, existing schemes for encoding assembly code often result in losing critical information, 
such as the parameters to call instructions, strings, and variable names. In particular, 
schemes like jTrans~\cite{wangJTransJumpawareTransformer2022} and PalmTree~\cite{liPalmTreeLearningAssembly2021} tend to normalize character text, constant numerical values, and external functions using special vocabulary tokens, which inadvertently leads to the omission of vital details. Second, models trained on  face difficulties when transferring the learned knowledge to different tasks, as they require a large amount of data for supervision, which is often downstream-specific and scarce in binary code analysis scenarios. 
\fi
% For instance, models like COMBO~\cite{zhang2022combo}, fine-tuned on the POJ-dataset~\cite{mou2016convolutional} for algorithm tasks, or jTrans~\cite{wangJTransJumpawareTransformer2022}, fine-tuned on BinaryCorp for binary code similarity detection (BCSD) tasks, rely on significant amounts of downstream-specific data for effective supervision.

% Transferability is essential in binary analysis due to the high cost and complexity of data collection in this field. 
Drawing inspiration from multi-modal learning, we observe that models such as CLIP~\cite{radfordLearningTransferableVisual2021} could learn better image representations by aligning visual modality concepts with human-comprehensible natural language. 
% This insight has inspired us to conceptualize assembly code as an analogous modality.
Similarly, we could conceptualize binary code as an analogous modality and explore the alignment between binary code and natural language, to develop more semantically profound representations of assembly code with better transferability.
% potentially yield a more powerful assembly representation learning model.
% 
% Proposed Solution and Challenges
Specifically, we could use natural language (i.e., explanations of code semantic) as a supervision signal for learning representations of binary code (i.e., assembly code) by aligning them with pre-training. The resulting model is highly likely to obtain representations that encapsulate more semantic information about binary code. 
\textit{To this end, we must address the following challenges}: (1) obtaining cost-effective aligned data for training and (2) preserving semantic information (e.g., control-flow and data-flow) extracted by existing binary code representation methods.

% Addressing the Challenges
To address these challenges, 
we introduce a novel method named \sysname, which utilizes natural language supervision to learn transferable binary code representations. 
Specifically, regarding the first challenge, we present an efficient dataset engine that automatically generates a large and diverse dataset of assembly code and natural language explanations for model pre-training. Regarding the second challenge, we introduce a novel binary code representation learning backbone network that combines WordPiece tokenization (e.g. used by Llama~\cite{touvronLLaMAOpenEfficient}) and jump-aware embedding (e.g., used by jTrans~\cite{wangJTransJumpawareTransformer2022}) designs. This network is capable of capturing both data flow and control flow information in binary code while preserving crucial information, such as function-call parameters, decompiler-processed strings, and variable names.

We have generated \textbf{195 million} pairs of aligned data and pre-trained a prototype model of \sysname, and evaluated it on several downstream tasks, including binary code similarity detection (BCSD), crypto-related functions identification, and protocol categorization. The results indicate that \sysname outperforms existing state-of-the-art (SOTA) solutions on these tasks, even without any further task-specific training or fine-tuning.
In the BCSD task, when searching for functions within 10,000 functions, \sysname without fine-tuning achieves the highest ranking with an average  Recall@1 rate of 83.3\%. In comparison, the current leading supervised solution only achieves a rate of 57.1\%. Regarding crypto-related function identification, \sysname performs slightly below the baseline in the zero-shot setting, but the fine-tuned version of \sysname largely surpasses the baseline by 17\%. In the protocol categorization experiment, both the zero-shot and finetuned versions of \sysname surpass the best baseline by 6\% and 23\%, respectively.
Furthermore, we conducted a case study to evaluate \sysname on real-world samples. The results demonstrate that \sysname has a remarkable application potential.
% Contributions

This paper principally explores a critical research question: \textit{Can a model be trained in the field of binary analysis to effectively transfer acquired knowledge to various tasks, even when confronted with extremely limited or non-existent data?} We affirmatively respond with \sysname, our innovative methodology that bridges the gap between binary code and natural language representations.
The principal contributions of our research are as follows:
\begin{enumerate}
 \item We introduce \sysname, which innovatively uses natural language as a supervisory signal to learn binary code representations. This method aligns assembly code with natural language and boosts transfer learning capabilities, especially in few-shot and zero-shot learning scenarios.

 \item We develop an efficient and scalable dataset engine capable of generating a comprehensive dataset of assembly code and corresponding natural language explanations. This dataset is instrumental in training models to represent assembly code with natural language supervision, bridging a significant gap in existing binary analysis methodologies.

 \item We conduct extensive experiments to demonstrate the effectiveness of our proposed \sysname method, which outperforms best baselines and showing remarkable transfer learning capabilities on various tasks. This showcases the model's versatility and effectiveness in binary analysis applications, setting a new benchmark in the field.

 \item We release our code and \sysname model at \url{https://github.com/Hustcw/CLAP} to the research community to facilitate future research.
\end{enumerate}

\section{Background and Related Works}\label{sec:background}

This section provides an overview of the key concepts and techniques relevant to \sysname, focusing on binary code representation and large language models.

\subsection{Binary Code Representation}

During the compilation process, high-level languages are transformed into assembly code (i.e. binary code), which is closer to machine hardware. Assembly code is more difficult to interpret compared to the source code, primarily because it lacks the clear abstractions present at the source code level, such as variable names and high-level logical structures. An example of assembly code is shown in Figure~\ref{fig:bubblesort}, which implements a bubble sort algorithm. Often in binary program analysis, there is a significant lack of access to the source code. This not only makes understanding the program more challenging but also spawns scenarios akin to the representation of the source code, underscoring the demand for effective representations of the assembly code.

Binary code representations are essential for various tasks such as binary code similarity detection (BCSD), function prototype inference, malware classification, and reverse engineering. 
Over the years, researchers have devised numerous techniques to represent binary code as continuous vectors in a vector space, making it suitable for downstream tasks. 
These methods for obtaining function vector representations can be broadly categorized into three groups: 
1) direct modeling of raw bytes, 
2) employing graph models to establish control flow relationships, and 
3) representing function instructions as instruction sequences.

\subsubsection{Raw Bytes}

Several studies, such as MalConv~\cite{raff2017malware}, DeepVSA~\cite{guo2019deepvsa} and $\alpha$-Diff~\cite{liu2018alphadiff}, 
% and research by Shin et al.~\cite{shinRecognizingFunctionsBinaries}, 
employ neural networks like CNNs~\cite{lecun1998gradient} and LSTMs\cite{hochreiter1997long} to analyze binary code from raw bytes, aiming to improve malware detection and identify code similarities. These approaches focus on capturing data dependencies and feature extraction without delving into instruction-level semantics or control flow graph structures, offering efficient computational processing but missing out on deeper semantic understanding.

% Several works utilize raw byte sequences for modeling binary code, treating a function as a flat, continuous sequence of raw bytes without considering instruction semantics or control flow information. MalConv~\cite{raff2017malware} parses executable files into byte sequences to analyze them with a neural network. On the other hand, DeepVSA~\cite{guo2019deepvsa} introduces a novel neural network structure that can capture data dependencies among instructions by utilizing LSTM~\cite{hochreiter1997long} on byte sequences. By doing so, the analysis of value sets can be remarkably enhanced. Additionally, $\alpha$-Diff~\cite{liu2018alphadiff} employs three semantic features to identify similarities between cross-version binaries. To extract internal semantic features from raw bytes of functions at the function level, it uses CNN~\cite{lecun1998gradient}. Furthermore, it extracts Inter-Func features between functions throughout call graphs and introduces Inter-Mod features by importing functions. Similarly, the work by Shin et al.~\cite{shinRecognizingFunctionsBinaries} employs LSTMs to learn from raw bytes.
% Although these methods are computationally efficient, they commonly lack the ability to grasp higher-level semantic information about each instruction and to consider CFG structural information.

\subsubsection{Graph Modeling}

Assembly code, inherently structured as a series of basic blocks connected by Control Flow Graphs (CFGs), has prompted studies to model CFGs graphically. Gemini~\cite{xu2017neural} utilizes GNNs to derive function embeddings from CFGs, though it is limited by the manual selection of features potentially missing complex semantics. GMN~\cite{li2019graph} innovatively calculates graph similarity via an attention-based Graph Matching Network. Both GraphEmb~\cite{cochardInvestigatingGraphEmbedding2022} and OrderMatters~\cite{vinyalsOrderMattersSequence2016} infuse DNNs to learn basic block attributes before applying graph models to encapsulate the interrelations of these blocks within the CFG framework, capturing the flow and structure more effectively.

% Graph-based approaches use graph neural networks (GNNs) to embed control flow information of basic blocks into assembly code embeddings. This results in a better understanding of functions by incorporating control flow information in function embeddings. Gemini~\cite{xu2017neural} generates function embeddings by utilizing a graph embedding model that is based on graph neural networks (GNNs), which are created from the control flow graph of each binary function. However, this process entails relying on manually-selected features to represent control flow graph (CFG) blocks. This can be problematic as it may not convey sufficient semantic information. Alternatively, GMN~\cite{li2019graph} proposes a novel Graph Matching Network that computes a similarity score between a pair of graphs by joint reasoning on the pair through a new cross-graph attention-based matching mechanism. GraphEmb~\cite{cochardInvestigatingGraphEmbedding2022} and OrderMatters~\cite{vinyalsOrderMattersSequence2016} both utilize Deep Neural Networks (DNN) to learn embeddings for each basic block, with the embeddings representing the attributes of basic blocks. Following this, a graph embedding model is applied to learn the Attributed Control Flow Graph (ACFG) embeddings, which capture the control flow graph. By doing so, these methods can effectively represent the relationships among basic blocks in the control flow graph.

\subsubsection{Sequence Modeling}

Sequence modeling methods consider assembly code as instruction series, capturing the instruction order. Asm2Vec~\cite{dingAsm2VecBoostingStatic2019} and SAFE~\cite{massarelliSAFESelfAttentiveFunction2019} use language-inspired models to generate embeddings for instructions and functions, treating instructions analogously to words. 
% Methods like INNEREYE~\cite{zuo2018neural} and RLZ2019~\cite{redmond2018cross} employ word2vec and LSTMs for learning from these instruction 'words'. 
Advanced techniques such as Trex~\cite{peiTrexLearningExecution2021} with Transformers, and jTrans'~\cite{wangJTransJumpawareTransformer2022} jump-aware model enhance micro trace and control flow representation, respectively. VulHawk~\cite{luoVulHawkCrossarchitectureVulnerability} integrates RoBERTa~\cite{liuRoBERTaRobustlyOptimized2019} and GCNs~\cite{kipf2016semi} for multi-faceted embeddings.
% , while BinBert~\cite{artusoBinBertBinaryCode2022} and Asteria~\cite{yangAsteriaProEnhancingDeepLearning} explore matching and semantic encoding. BinShot~\cite{ahn2022practical} further refines this approach using a BERT-based Siamese model for distance vector learning. 
Despite their effectiveness in capturing semantic details, most sequence models lack comprehensive control flow graph representation, with some like jTrans attempting to include this aspect but it loses disassembled string information and external function call information during tokenization and normalization.

\begin{figure}[t!]
    \centering
    \setlength{\abovecaptionskip}{2mm}
    \includegraphics[width=\linewidth]{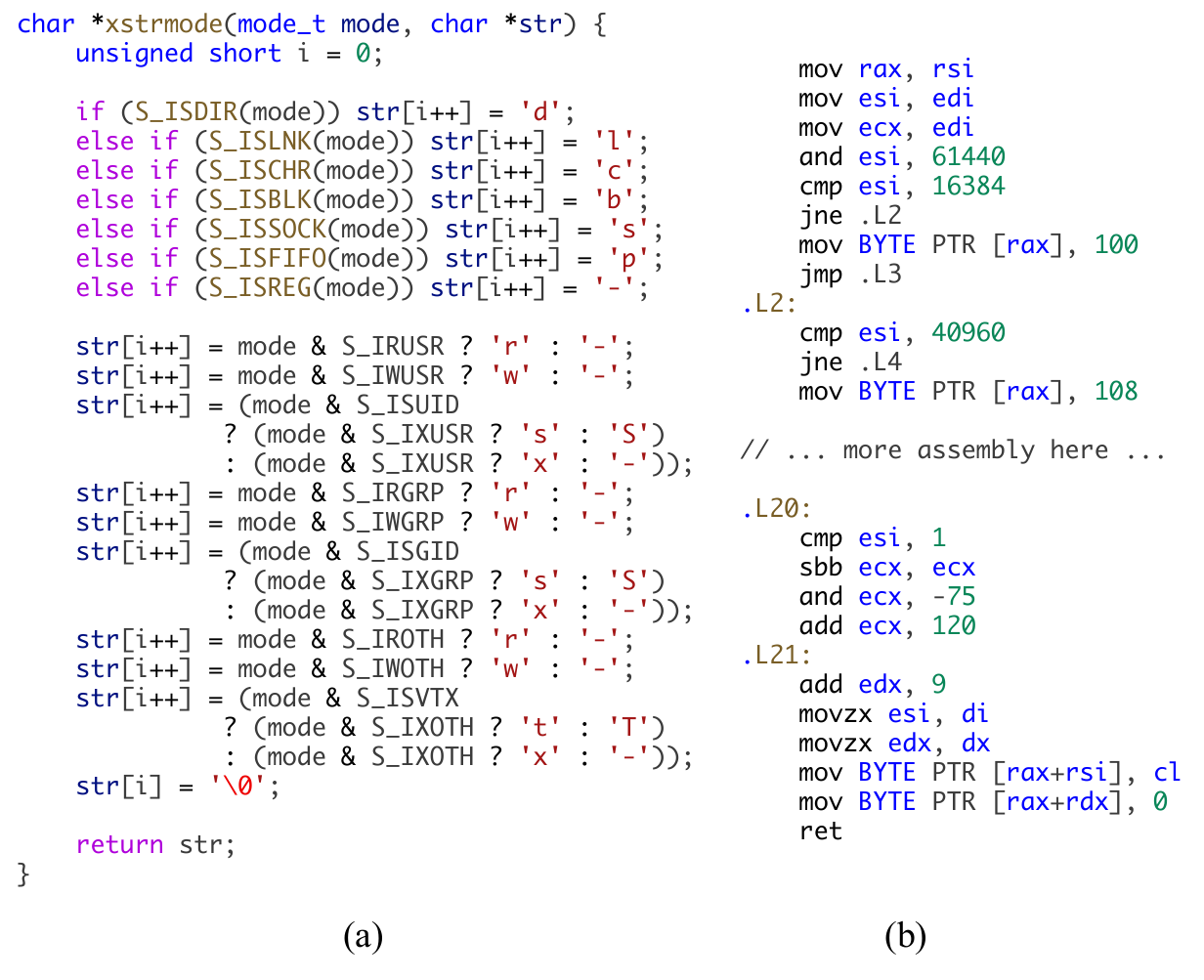}
    \caption{A real-world function named \texttt{xstrmode} (a) and its assembly code (b)}
    \label{fig:xstrmode}
    \vspace{-2mm}
\end{figure}

\begin{figure*}[t!]
  \centering
    \setlength{\abovecaptionskip}{2mm}
  \includegraphics[width=1\linewidth]{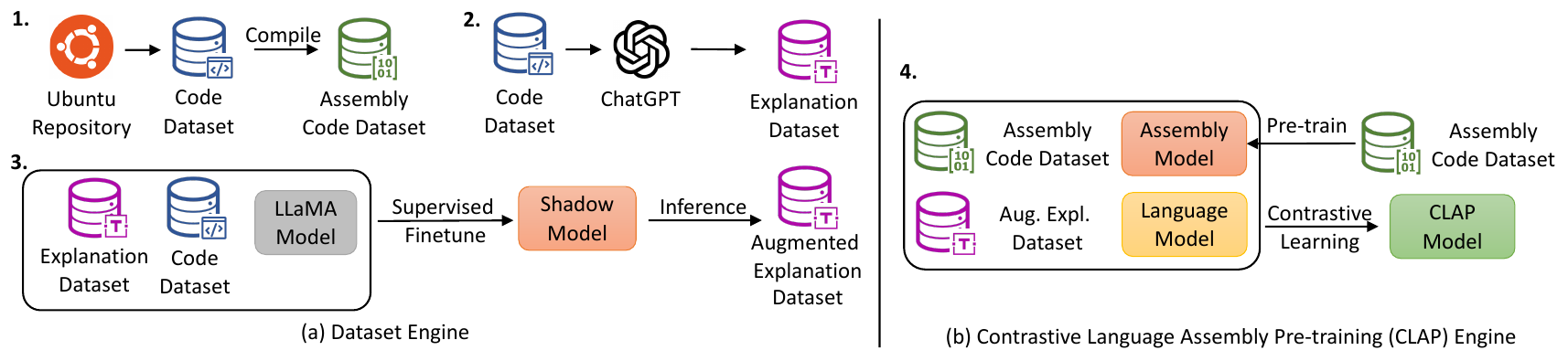}
  \caption{
  Overview of our primary workflow, consisting of two main components, the dataset engine and the \sysname engine. 
  The dataset engine compiles source code from the Ubuntu Repository into assembly code and uses GPT to generate explanations, forming an explanation dataset. We then fine-tune LLaMA model using the source code with the corresponding explanation, resulting in a shadow model that generates the augmented explanation dataset. Within the \sysname engine, we first pre-train the assembly code dataset to develop an initial assembly encoder. By employing contrastive learning with this assembly encoder and a text encoder on both assembly code and augmented explanation dataset, we produce the final \sysname model.}
  \label{fig:workflow}
\end{figure*}

\subsection{Large Language Models (LLM)}

Recently, large language models especially the GPT series~\cite{borgeaudImprovingLanguageModels2021,radfordLanguageModelsAre,brownLanguageModelsAre2020} of models, such as GPT-3.5 (i.e. ChatGPT)~\cite{gpt-3.5} and GPT-4~\cite{openaiGPT4TechnicalReport}, have demonstrated extraordinary source code understanding capabilities. 
For instance, we prompt ChatGPT to output the explanation the code snippet shown in Figure~\ref{fig:xstrmode}a, ChatGPT demonstrates its robust comprehension of source code not only through practical usage but also in its ability to elucidate code functions, i.e. ``The purpose of this code is to create a string that represents the mode of a file or directory like -rw-r--r--''.

In contrast, these models exhibit less proficiency with assembly code. When tasked to explain the assembly code for \texttt{xstrmode} shown in Figure~\ref{fig:xstrmode}b, 
% the explanation provided by 
ChatGPT is largely ineffective. This can be attributed to the challenges in capturing control flow information and the scarcity of assembly code in pre-training corpora.

\section{Methodology}
\label{sec:method}

Figure~\ref{fig:workflow} presents the core workflow of our study, detailing the essential stages of our process. 
Our methodology begins with extensive compilation activities on a 96-core server using the Ubuntu repository~\cite{ubuntu}, which lasted three months. This process produced a diverse range of assembly codes, generated via various optimization techniques and compilers while retaining their corresponding source codes. For initial interpretations at the source code level, we utilized GPT-3.5, employing tailored prompts to bootstrap our data. Subsequently, we fine-tuned the LLaMA model~\cite{touvronLLaMAOpenEfficient} with this curated dataset. This fine-tuning was aimed at customizing the model’s output to our specific requirements and performing local inference, which resulted in an expanded set of source code explanations for data augmentation.

% Figure~\ref{fig:workflow} illustrates the primary workflow presented in this paper, providing an overview of the key processes involved.
% We perform numerous compilations using the Ubuntu repository~\cite{ubuntu} on a 96-core server lasting for three months, resulting in assembly codes generated with various optimizations and compilers while preserving the corresponding source codes. We acquire source code-level interpretations from GPT-3.5 as our bootstrap using appropriate prompts. We then fine-tune LLaMA~\cite{touvronLLaMAOpenEfficient} locally with the obtained data to align its output format with our needs and conducted local inference, yielding additional source code explanations for data augmentation.

To overcome the challenges outlined in Section~\ref{sec:intro}, we introduce an innovative assembly encoder model, meticulously designed to capture both the semantic and structural nuances of assembly code. Initially, we pre-trained a jump-aware transformer on an extensive assembly code dataset, aiming to create an expressive and effective encoder model. Then, employing a contrastive learning strategy with a multitude of negative samples, we effectively aligned this assembly code encoder with a text encoder, which itself was pre-trained on more than 1 billion text pairs~\cite{all-mpnet-base-v2}. Through the optimization of the InfoNCE loss, we significantly enhanced the mutual information between the positive and negative samples of assembly code and explanations. This optimization facilitated the precise matching of assembly code with text explanations and the successful alignment of their embeddings.

\subsection{Dataset Engine}

In this section, we will discuss the process of obtaining the training dataset. The core idea is to utilize LLMs' understanding of source code and construct a relationship between source code and assembly code to obtain natural language explanations for assembly code. Considering the cost of building the dataset, apart from using GPT, we locally trained a model capable of generating sufficiently good natural language explanations to create a large and diverse dataset.

\subsubsection{Binary Generation}
\label{subsubsec:binary-generation}

We utilize package managers to achieve large-scale automated compilation and source code acquisition, specifically using Ubuntu's package manager. Ubuntu is one of the most popular distributions, containing many C/C++ packages, each with a script that compiles the source code into the final product. This makes it suitable for generating a diverse set of binary executable files along with their source code.
% When compiling the source code for Ubuntu's package manager, we replaced the default system compiler with our custom compiler forwarder and adjusted the compilation options to meet our requirements, such as altering optimization options within the compilation settings or employing \texttt{Clang} for compilation, even if the package manager's tool invoked \texttt{GCC}. This method proved to be more efficient and rigorous than merely configuring environment variables, such as \texttt{CFLAGS} and \texttt{CC}, to achieve diverse optimization levels and compilers.

We employ a variety of compilers and optimization levels to augment the quantity of assembly code, which contains six different compilers (\texttt{GCC-\{7,9,11\}} and \texttt{Clang-\{9,11,12\}}) and five optimization levels (\texttt{O\{0-3\}} and \texttt{Os}). Additionally, we maintain both stripped and non-stripped functions to facilitate assembly pre-training and contrastive pre-training in later stages.

% In order to obtain more accurate source code correspondence compared to directly obtain it from the original file and to retain more semantic information in variable names and macro definition names compared to decompiling the assembly code to aid source code understanding and interpretation for LLM, we modify the compiler's behavior during compilation. Specifically, we instruct the compiler to omit directives like \texttt{ifdef} and \texttt{elseif}, maintain comments, and avoid macro expansion by adding \texttt{-E} option during compilation. Furthermore, when obtaining corresponding source code from the final binary's assembly code, it employs debug information to identify the actual source code addresses for each function in the binary, reducing problems caused by identically named functions within the same project.

We utilize Clang~\cite{clang} to parse the preprocessed source code, extracting source code snippets by determining the position of each function within the code file. To enhance the semantic value of the functions and minimize the noise in the dataset, we exclude source code segments with less than three lines and eliminate assembly code containing less than three basic blocks.

\subsubsection{Source Code Explanation}

We employ GPT-3.5 to generate natural language explanations for source code, as it lacks comprehension of assembly code. By leveraging the correspondence between source code and assembly code, we aim to create sufficiently detailed text explanations that can explain the source code and assembly code at the same time.
% During the design process of the prompt for LLM, we ensure the model focuses on capturing an overview of the code's execution while avoiding detailed step-by-step explanations, which are often difficult to obtain at the assembly level. This is because, during testing, we observed that the model tended to naturally perform CoT~\cite{weiChainThoughtPrompting2022} and yield verbose explanation. Additionally, we prevent output that includes variable names or other information not present in the stripped binary. At last, we let the model generate multiple tags to categorize the code segment.

% \wh{minimize the principles}In the design of prompts, we adhere to several principles: 1. We aim for the model to capture a high-level summary of code execution, eschewing granular, step-by-step descriptions. The rationale is that, in our testing phase, we noticed the model is inclined toward verbose explanations through Chain of Thought (CoT)~\cite{weiChainThoughtPrompting2022}, which becomes impractical at the assembly level. 2. Information loss inherent in code compilation results in missing details, like variable and structure names, in assembly language that are irretrievable. Consequently, we guide the model to exclude such details from its output. 3. We configure the model to produce a variety of tags that serve to classify segments of code efficiently. We show the prompt used to request GPT-3.5 in Table~\ref{tab:explanation-prompt} and conduct the evaluation on the quality of explanation in Section~\ref{subsec:data-engine-evaluation}.

We manually design the instructions for GPT-3.5 to generate an explanation. We first prompt the model to concisely summarize the source code while avoiding detailed procedural narrations.
Secondly, acknowledging the intrinsic information loss during code compilation, such as the absence of variable names, we instruct the model to forgo these non-recoverable specifics. Lastly, we program the model to assign tags to the code snippet. 
% We show the prompt used to request GPT-3.5 in Table~\ref{tab:explanation-prompt} and conduct the evaluation on the quality of explanation in Section~\ref{subsec:data-engine-evaluation}.
% \elsa{todo: Add prompt in github}

\iffalse
\begin{table}[t!]
\centering
\fontsize{10.0pt}{\baselineskip}\selectfont
\linespread{1.0}\selectfont
\begin{boxedminipage}{1.0\columnwidth}
% In-context example 1
\mybox{
\small
Explain the purpose of each function in the given code snippet in a simple, non-technical way. Focus on real-world applications and avoid using technical terms or code details. Keep your response within 150 words, so it's easy to understand for someone without coding knowledge. Connect the explanation to everyday concepts, emphasizing the overall meaning and functionality of the code. In the end, assign tags for the code, and keep any other text to a minimum.
}

\end{boxedminipage}
\caption{
Prompt used for source code explanation.
}
\label{tab:explanation-prompt}
\end{table}
\fi

% \elsa{Add how we do the filtering on source code and asm}

\subsubsection{Shadow Model}

% We conduct supervised finetuning (SFT) on the LLaMA model using half of the 2.6 million function explanations from GPT-3.5, initializing it with the aligned Vicuña 13B model~\cite{vicuna2023} and performing complete finetuning to create a shadow model. For comparison, we train two additional models, one aligning LLaMA 13B and another aligning LLaMA 30B, both using 50,000 function explanations. We will evaluate and compare their performances in the Section~\ref{subsec:data-engine-evaluation}.

We conduct supervised finetuning (SFT) on the Vicuña 13B~\cite{vicuna2023} using half of the 2.6 million function explanations from GPT-3.5. 
% The finetuning process involves using an Adam optimizer with a learning rate of 1e-5, a cosine learning rate scheduler and warmup ratio of 0.03, batch size of 128. 
For comparison, we train two LLaMA 13B and 30B models respectively, using 50,000 explanations of source code function. We evaluate and compare their performances in Section~\ref{subsec:data-engine-evaluation}.
After obtaining the shadow model, we employ it to facilitate the generation of a larger corpus of natural language explanations. This method serves as a data augmentation technique, enabling us to produce an expanded dataset of natural language explanations and assembly code pairs. As a result, the enriched dataset significantly increases the quantity of available data for Section~\ref{subsubsec:contrastive-learning} while maintaining acceptable quality.

\begin{figure*}[!t]
  \centering
  \setlength{\abovecaptionskip}{2mm}
  \includegraphics[width=0.8\linewidth]{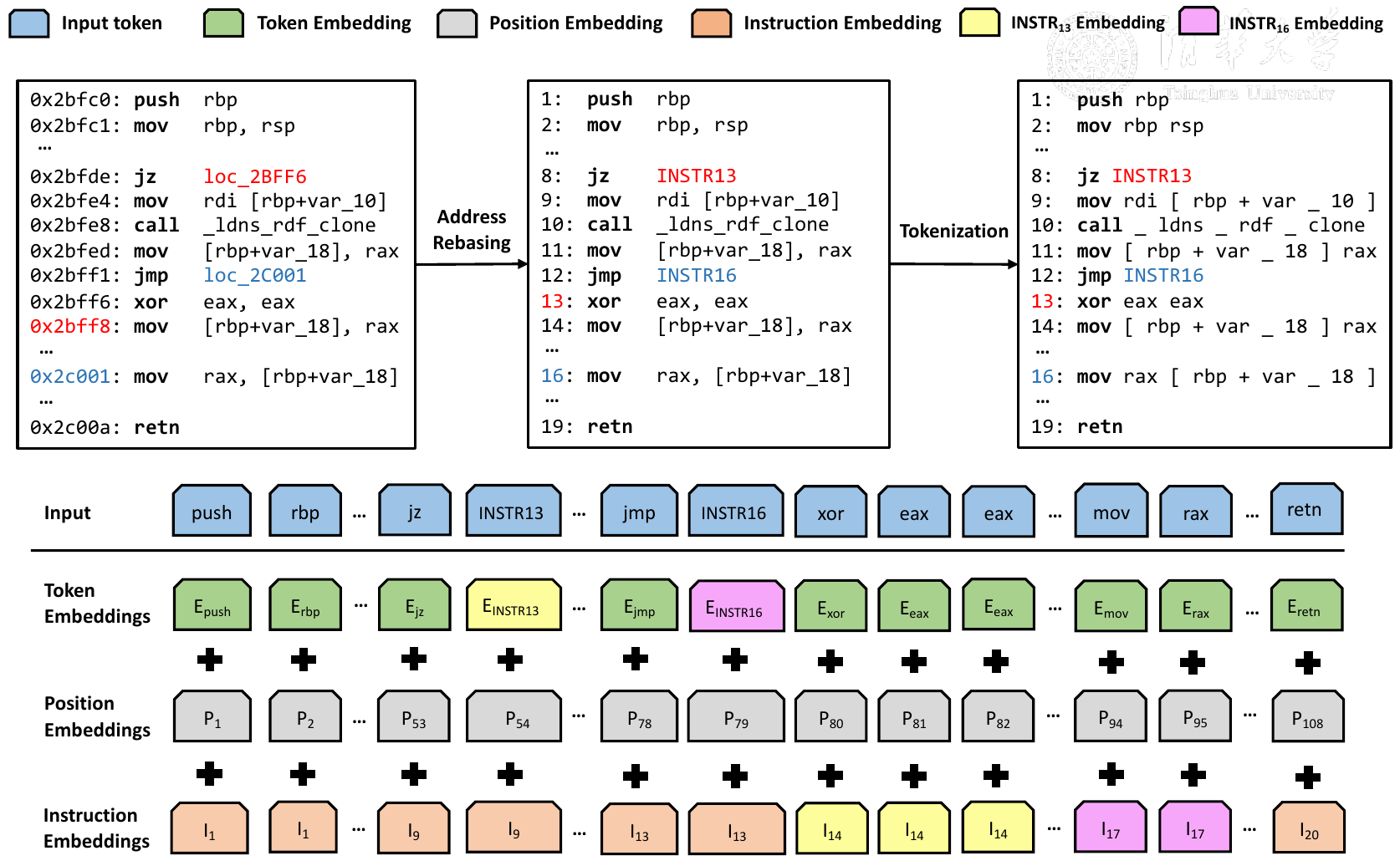}
  \caption{Illustration of \asmencoder. The raw assembly code is first rebased and tokenized. Then, each token is converted to a {\em token embedding}, a {\em position embedding}, and an {\em instruction embedding}, and the sum of these three is output as the embedding. The instruction embedding remembers instruction boundaries and works with jump symbols to comprehend control flow. The token embedding of the jump symbol (e.g., $\mathbf{E_\mathtt{INSTR13}}$) shares parameters with the target instruction embedding (e.g., $\mathbf{I_{\mathtt{13}}})$.}
  \label{fig:asm-encoder}
\end{figure*}

\subsection{Contrastive Language Assembly Pre-training (\sysname) Engine}

Next, we will introduce the design and training methodology of the entire \sysname engine, including the design of the assembly encode (i.e., \asmencoder), the training process of the \sysname engine, and the rationale behind using natural language as a supervisory signal to enhance transferability. 

\subsubsection{Model Architecture}

\asmencoder is based on the RoBERTa base architecture~\cite{liuRoBERTaRobustlyOptimized2019} with 110M parameters. We average the output of the last Transformer layer as the assembly embedding. As assembly code is quite different from natural languages. To effectively represent the assembly code while capturing its inherent structure and semantics while preserving important information such as function-call parameters, decompiler-processed strings, and variable names. The \asmencoder is illustrated in Figure~\ref{fig:asm-encoder}, we have adopted the following strategies in designing the assembly encoder.

\paragraph{Instruction Embedding}We incorporate instruction embedding into the Transformer model to delineate the boundaries of assembly instructions. This allows the model to identify and process individual instructions effectively. By assigning unique embeddings to each instruction, the model learns to differentiate between various instructions. This distinction is crucial for the next procedures.

\paragraph{Address Rebasing} As shown in Figure~\ref{fig:asm-encoder}, we first rebase the address of the assembly code, and we preserve the relative address relationships while handling jump instructions. Instead of normalizing different addresses to the same token like in previous work, this approach retains the control flow information within the assembly code. Address rebasing keeps the relative distance between jump instructions and their targets, enabling the model to understand the control flow changes caused by jump instructions.

\paragraph{Tokenization} After address rebasing, we tokenized the assembly code to separate tokens. We first employ the WordPiece~\cite{kudo2018subword} algorithm to train a tokenizer on the whole assembly code datasets, which is specifically tailored for assembly code. This tokenizer can perform a lossless encoding of assembly code without normalization, thus preserving crucial information such as calling parameters and external function names. In this way, we enable the model to handle the diverse vocabulary of assembly languages, including opcodes, registers, constants, and literals. 

\paragraph{Jump Relationships} We enhance the Transformer's understanding of jump relationships within the assembly code by sharing the token embedding parameters of jump symbols and their corresponding instruction embeddings as shown in Figure~\ref{fig:asm-encoder}. For example, the token embedding $\mathbf{E_\mathtt{INSTR16}}$ shares parameters with the instruction embedding $\mathbf{\mathtt{I_{16}}}$. This shared representation aids the model in capturing the complex connections between various components of the assembly code. By sharing the parameters, the model learns to associate symbol tokens with their respective target tokens, allowing it to better understand the control flow within the code. 

\subsubsection{Assembly Encoder Pre-training}
We first pre-train the assembly encoder to comprehend assembly code by selectively masking tokens within the assembly context, which has been proven highly effective in previous work. We follow the same training task of previous work jTrans~\cite{wangJTransJumpawareTransformer2022}, including the Masked Language Model (MLM) task and the Jump Target Prediction (JTP) task.
The collective loss function $\mathcal{L}_{P}(\theta)$ in the pre-training phase is the combination of MLM and JTP objective functions in Equation~\ref{eq:mlm-pretrain-loss}, where $x_{i}$ is the $i$-th token, $\mathbf{m_{x}}$ and $\mathbf{l_{x}}$ denotes for the masking positions for normal tokens and jump tokens.

\begin{equation} 
% \small \min_{\theta} \mathcal{L}_{P}(\theta) = \mathcal{L}_{MLM}(\theta) + \mathcal{L}_{JTP}(\theta) 
\small \min_{\theta} \mathcal{L}_{P}(\theta) = \sum_{i \in \mathbf{m_{x}}} -\log P(x_i|\mathbf{f}^{\text{~mlm}}) + \sum_{i \in \mathbf{l_{x}}} -\log P(x_i|\mathbf{f}^{\text{~jtp}})
\label{eq:mlm-pretrain-loss}
\end{equation}

% To solve the challenge mentioned in Section~\ref{sec:intro}, we adopt BPE tokenizer to refrain from normalization of strings and numerals, thus circumventing issues related to Out-of-Vocabulary (OOV). Moreover, the learnable position embedding~\cite{vaswaniAttentionAllYou2017} of jTrans is replaced by a rotary position embedding (RoPE)~\cite{suRoFormerEnhancedTransformer2022} to enhance performance. 
Through the first stage of pre-training, we obtain a well-initialized set of weights for the second stage of contrastive pre-training, facilitating a hastened convergence of the contrastive pre-training. 

\begin{figure}[t!]
    \centering
    \setlength{\abovecaptionskip}{2mm}
    \includegraphics[width=1\linewidth]{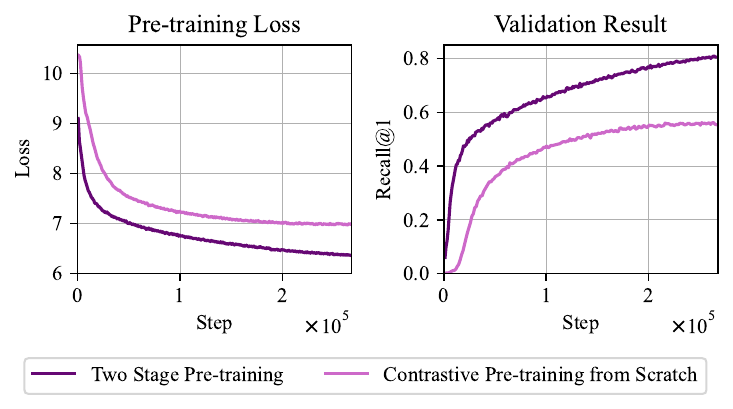}
    \caption{The comparison between contrastive pre-training from scratch and contrastive pre-training based on the first pre-training stage model (batch size = 1024, epoch <= 1). The left figure shows the InfoNCE loss during pre-training. The right figure shows the Recall@1 result in the validation dataset, in which the model needs to select the sole natural language explanation that matches the assembly code from 65,536 natural language explanations.}
    \label{fig:compare-from-scratch-pretrained}
    \vspace{-2mm}
\end{figure}

\subsubsection{Contrastive Pre-training with Natural Language Supervision}
\label{subsubsec:contrastive-learning}
After pre-training, we utilize a contrastive learning approach to align the assembly code encoder with a text encoder pre-trained on a large collection of text pairs~\cite{all-mpnet-base-v2}. The key to implementing natural language supervision with contrastive learning lies in using the natural language representations obtained from pre-trained natural language models as anchors. The training task is to predict which explanation corresponds to which assembly code.
We apply the InfoNCE loss~\cite{oord2018representation} to discriminate assembly code with the positive text explanation and negative text explanations. Given a positive pair assembly code and text explanation $(a_{i}, t_{i})$ and a set of negative pairs ${(a_{i}, t_{j})}_{j \neq i}$. We obtain representations for assembly code and explanations through assembly encoder $\mathcal{F}_{A}$ and text encoder $\mathcal{F}_{T}$, respectively,

\begin{equation}
\small
% \begin{aligned}
 E_{A_i} = \mathcal{F}_{A}(a_i),\quad
 E_{T_i} = \mathcal{F}_{T}(t_i)
% \end{aligned}
% \label{eq:contrastive-pretrain-loss}
\end{equation}

where the assembly embedding for assembly code $A_{i}$ denotes for $E_{A_i}$, text embedding for explanation $T_{j}$ denotes for $E_{T_i}$ and $N$ denotes for the number of samples in a single batch.
The InfoNCE loss is defined as Equation~\ref{eq:contrastive-pretrain-loss}

\begin{equation}
\small \mathcal{L}_{E} = - \log \frac{\exp(E_{A_{i}} \cdot E_{T_{i}})}{\sum_{j=1}^{N} \exp(E_{A_{i}} \cdot E_{T_{j}})}
\label{eq:contrastive-pretrain-loss}
\end{equation}

By optimizing the InfoNCE loss, we enhance the mutual information between assembly code and explanations, effectively aligning their representations. In an implementation, we employ a well-pre-trained text encoder $\mathcal{F}_{T}$ and utilize an exceptionally large batch size $N=65,536$. This approach enables our model to accurately identify the most fitting explanation from a vast pool of options for each assembly code. It allows our model to leverage the rich semantics of natural language explanations to augment the learned assembly code representations.

We compare contrastive learning from scratch and contrastive learning based on the first-stage model. The contrastive pre-training loss and the validation result are illustrated in the Figure~\ref{fig:compare-from-scratch-pretrained}. The loss of the latter model decreases more rapidly during training and it achieves higher validation results. Compared to contrastive pre-training, the first pre-training stage incurs less than 10\% of the cost, highlighting the efficiency of the two-stage pre-training.

% With the \sysname method, our model is capable of achieving remarkable zero-shot learning capabilities across various tasks, such as binary code similarity detection and protocol categorization, which are evaluated in Section~\ref{subsec:zero-few-shot}. This demonstrates the effectiveness of our approach in utilizing text supervision to guide assembly code representation learning and improve transfer learning capabilities.

\subsection{Zero-Shot Inference Capability}
\label{subsec:zero-shot-inference-for-downstream-tasks}

\begin{figure}[t]
    \centering
    \setlength{\abovecaptionskip}{2mm}
    \includegraphics[width=0.95\linewidth]{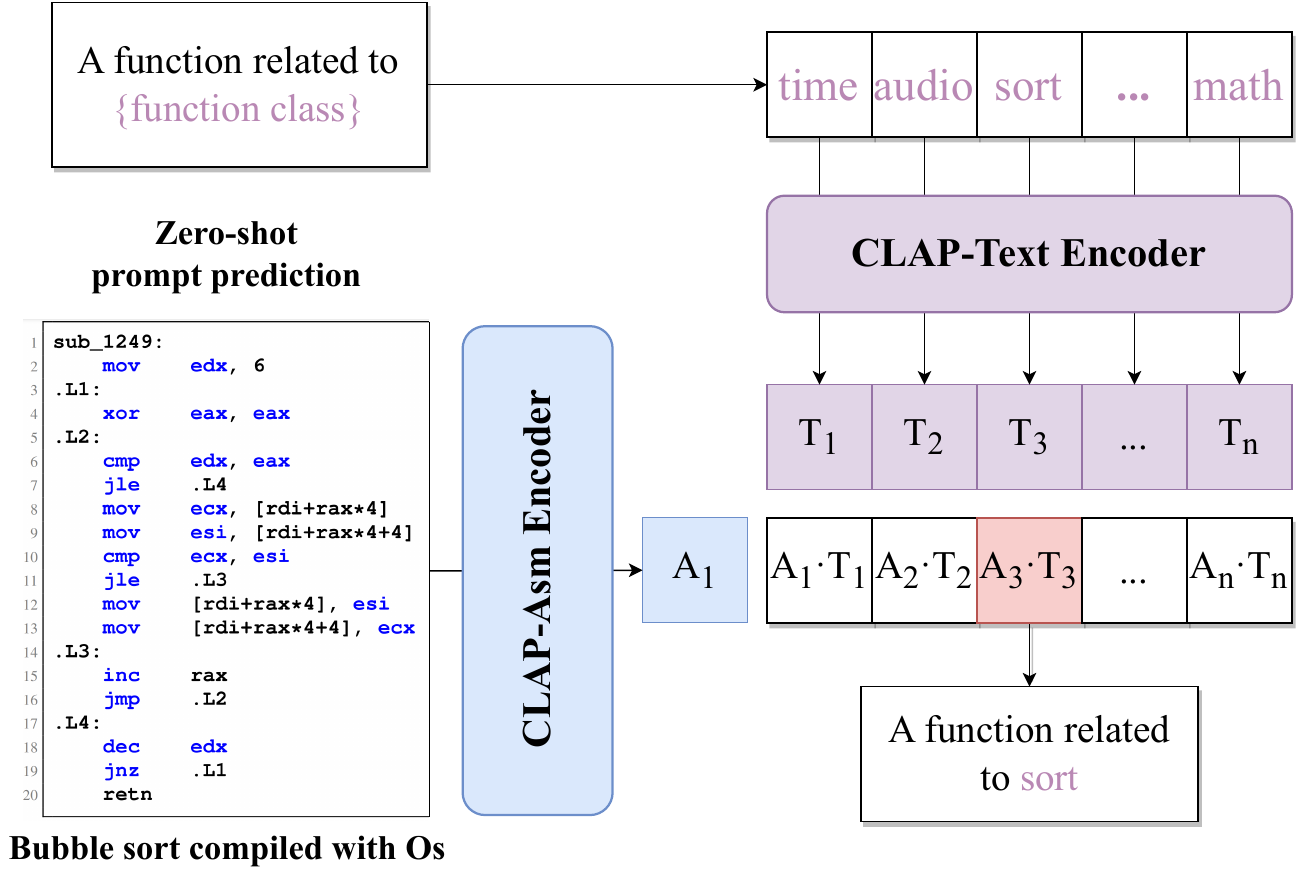}
    \caption{Zero-Shot Inference of \sysname}
    \label{fig:zero-shot-inference}
        \vspace{-2mm}
\end{figure}

As we introduce the rich semantic content of natural language into the representation of assembly code through contrastive learning, we can achieve impressive zero-shot capabilities in downstream assembly code understanding tasks. We demonstrate in the Figure~\ref{fig:zero-shot-inference} how we use zero-shot for downstream tasks of assembly code understanding.
% \wh{reduce this part} In a multi-class (such as n-class) classification scenario, we prepare $n$ labels in advance, such as `time', `sort', `audio', etc. First, we encode the assembly code with \sysname-ASM to get embedding \(A_1\), then for the natural language side, we design prompts that can be filled with predefined labels, such as `A function related to \{function class\}', where `\{function class\}' can be replaced by predefined labels, resulting in n sentences, which are then encoded with \sysname-TEXT to get $n$ embeddings \(T_i, i \in [1, n]\). Finally, we calculate the similarity of the assembly code embedding \(A_1\) and the \(i\)th natural language description embedding \(T_i\) by computing \(A_1 \cdot T_i\) to obtain the embedding most similar for zero-shot results.
In a multi-class classification task, the assembly code is first encoded into an embedding \(A_1\) using \asmencoder. Concurrently, n prompts are constructed to describe the multi-class, which are subsequently processed by \sysname-Text to generate n corresponding text embeddings \(T_i\). To derive zero-shot results, we calculate the dot product of $A_1$ and each $T_i$ to compute the similarity. The prompt with the highest similarity is marked as the classification result.
% embedding by calculating their dot product, pinpointing the one with the highest similarity.
With each similarity as logits, we can apply the SoftMax function on $[A_1 \cdot T_1, A_1 \cdot T_2, \cdots , A_1 \cdot T_n]$ to get the probability of each label.

% \elsa{add a argmax ... equation}

% On the left of Figure~\ref{fig:zero-shot-inference}, the assembly code is from a bubble sort function compiled using the Os optimization option, and among the predefined multiple categories, the embedding of prompt $T_3$ for the `sort' class, i.e. `A function related to sort', has the highest similarity with the embedding of the assembly code $A_1$, indicating that the zero shot classification result is that the function is related to sorting. In Section~\ref{subsec:sort-algorithm-classification}, we also show that the model has zero-shot discrimination capabilities for more refined concepts such as `Quick Sort' and `Heap Sort'.

\section{Evaluation}
We implement \sysname using Pytorch 2.0~\cite{paszke2019pytorch}. We used IDA Pro 7.6~\cite{idapro} to disassemble and extract the functions from the binary code in all of the experiments. Our training and experiments are conducted on several servers to
accelerate training. The CPU setup is 128 cores with 2TB RAM for each server. The total GPU setup is 32 NVIDIA Tesla A100.
We conduct extensive experiments to address the following questions: 

\begin{itemize}

\item \textbf{RQ1:} How is the quality of the assembly code representations learned by \sysname?

\item \textbf{RQ2:} Does the representation learned by \sysname exhibit strong transferability, even when confronted with extremely limited or non-existent data?

\item \textbf{RQ3:} What is the underlying theoretical reason why natural language explanations could support binary code representation learning?

\item \textbf{RQ4:} What is the quality of the dataset obtained by the dataset engine?

\end{itemize}

\subsection{Quality of \sysname Representations (RQ1)}

In representation learning, the efficacy of different models' representation is typically evaluated by freezing the pre-trained model and training only a linear projection layer for downstream task assessment, known as linear probe~\cite{liuCMACLIPCrossModalityAttention2021}. In our evaluation, we scrutinize a range of downstream tasks, notably the well-studied BCSD task~\cite{cochardInvestigatingGraphEmbedding2022,vinyalsOrderMattersSequence2016,peiTrexLearningExecution2021,wangJTransJumpawareTransformer2022}, along with various classification tasks. This methodology aims to confirm that the \sysname Model has effectively acquired a more sophisticated representation of assembly code.

% In this evaluation, we assess multiple downstream tasks, including the frequently explored BCSD task~\cite{cochardInvestigatingGraphEmbedding2022,vinyalsOrderMattersSequence2016,peiTrexLearningExecution2021,wangJTransJumpawareTransformer2022}, and several classification tasks to illustrate that the \sysname Model has successfully learned an improved representation of assembly code.

\subsubsection{Binary Code Similarity Detection (BCSD)}

% \elsa{todo: Add background of BCSD}
% Binary code similarity detection (BCSD), which identifies the degree of similarity between two binary codes, is a valuable tool for a wide range of applications, including the discovery of known vulnerabilities.

% BCSD is crucial in binary analysis for tasks such as identifying vulnerabilities and detecting code plagiarism in binary fragments.
% It's especially important for detecting vulnerabilities in software chains, where it's essential to quickly and accurately find affected software from binary code when source code isn't available. 
% BCSD is a fundamental metric in binary analysis, with extensive research efforts aimed at enhancing its effectiveness and precision for critical tasks in software security and maintenance.
BCSD represents a crucial approach to software security. It attempts to identify the degree of similarity between two assembly codes. This method is integral to binary analysis, with significant research dedicated to augmenting its accuracy and efficiency, essential for tasks such as vulnerability identification, malware detection, and supply chain analysis.
According to~\cite{marcelliHowMachineLearning2022}, deep learning-based methods~\cite{liGraphMatchingNetworks2019,yuCodeCMRCrossModalRetrieval2020,dingAsm2VecBoostingStatic2019,massarelliSAFESelfAttentiveFunction2019} now have been becoming the de facto SOTA for BCSD, outperforming the traditional methods.
We evaluate \sysname and several baselines using the \binarycorp~\cite{wangJTransJumpawareTransformer2022} datasets. The baselines include 
Gemini~\cite{xuNeuralNetworkbasedGraph2017}, 
GNN~\cite{liGraphMatchingNetworks2019}, 
GraphEmb~\cite{massarelliInvestigatingGraphEmbedding2019}, 
OrderMatters~\cite{yuOrderMattersSemanticAware2020}, 
SAFE~\cite{massarelliSAFESelfAttentiveFunction2019}, 
Asm2Vec~\cite{dingAsm2VecBoostingStatic2019}, 
Trex~\cite{peiTrexLearningExecution2021}, 
PalmTree~\cite{liPalmTreeLearningAssembly2021} and jTrans~\cite{wangJTransJumpawareTransformer2022}.

\begin{table*}
\centering
\caption{Comparison between \sysname and baselines for the BCSD task on \binarycorp-3M dataset (Poolsize=10,000)}
\scalebox{0.8}{
\begin{tabular}{c|ccccccc|ccccccc}
\Xhline{1pt}
\hline
             & \multicolumn{7}{c|}{\textbf{MRR}}                             & \multicolumn{7}{c}{\textbf{Recall@1}}                               \\ \hline
\textbf{Models}       & \textbf{O0,O3} & \textbf{O1,O3} & \textbf{O2,O3} & \textbf{O0,Os} & \textbf{O1,Os} & \textbf{O2,Os} & \textbf{Average} & \textbf{O0,O3} & \textbf{O1,O3} & \textbf{O2,O3} & \textbf{O0,Os} & \textbf{O1,Os} & \textbf{O2,Os} & \textbf{Average}  \\ \hline
Gemini         & 0.037 & 0.161 & 0.416 & 0.049 & 0.133 & 0.195 & 0.165   & 0.024 & 0.122 & 0.367 & 0.030 & 0.099 & 0.151 & 0.132    \\
GNN & 0.048 & 0.197 & 0.643 & 0.061 & 0.187 & 0.214 & 0.225 & 0.036 &  0.155 & 0.592 & 0.041 & 0.146 & 0.175 & 0.191 \\ 
OrderMatters & 0.062 & 0.319 & 0.600 & 0.075 & 0.260 & 0.233 & 0.263 & 0.040 & 0.248 & 0.535 & 0.040 & 0.178 & 0.158 & 0.200          \\ 
GraphEmb     & 0.087 & 0.217 & 0.486 & 0.110 & 0.195 & 0.222 & 0.219 & 0.050 & 0.154 & 0.447 & 0.063 & 0.135 & 0.166 & 0.169          \\
SAFE         & 0.127 & 0.345 & 0.643 & 0.147 & 0.321 & 0.377 & 0.320 & 0.068 & 0.247 & 0.575 & 0.079 & 0.221 & 0.283 & 0.246  \\
Asm2Vec      & 0.072 & 0.449 & 0.669 & 0.083 & 0.409 & 0.510 & 0.366 & 0.046 & 0.367 & 0.589 & 0.052 & 0.332 & 0.426 & 0.302  \\
PalmTree & 0.130 & 0.403 & 0.677 & 0.152 & 0.355 & 0.496 & 0.369 & 0.08 & 0.326 & 0.609 & 0.097 & 0.281 & 0.420 & 0.302 \\ 
Trex & 0.118 & 0.477 & 0.731 & 0.148 & 0.511 & 0.513 & 0.416 & 0.073 & 0.388 & 0.665 & 0.088 & 0.422 & 0.436 & 0.345 \\ 
jTrans (Zero Shot)       & 0.137 & 0.490 & 0.693 & 0.182 & 0.472 & 0.510 & 0.414 & 0.088 & 0.412 & 0.622 & 0.122 & 0.393 & 0.430 & 0.340 \\
jTrans (Linear Probe)  & 0.333 & 0.573 & 0.715 & 0.404 & 0.608 & 0.601 & 0.539 & 0.245 & 0.494 & 0.644 & 0.309 & 0.526 & 0.520 & 0.456 \\
jTrans (Finetune)  & 0.475 & 0.663 & 0.731 & 0.539 & 0.665 & 0.664 & 0.623 & 0.376 & 0.580 & 0.661 & 0.443 & 0.586 & 0.585 & 0.571 \\ \hline
\sysname (Zero shot) & \textbf{0.764} & \textbf{0.903} & \textbf{0.941} & \textbf{0.813} & \textbf{0.906} & \textbf{0.877} & \textbf{0.867} & \textbf{0.719} & \textbf{0.875} & \textbf{0.920} & \textbf{0.774} & \textbf{0.881} & \textbf{0.847} & \textbf{0.836} \\
\Xhline{1pt}
\end{tabular}
}\label{tab:binarycorp-10000}
\end{table*}

% Consistent with previous studies, we train jTrans, Trex and PalmTree on the training set of \binarycorp~\cite{wangJTransJumpawareTransformer2022} and evaluate their performance on the corresponding test set.
To assess the effectiveness of the embeddings generated by pre-trained models, we incorporate a Linear Probe~\cite{radfordLearningTransferableVisual2021} into their respective outputs, enabling the linear layer to project these outputs into an appropriate vector space. More specifically, we add a linear classifier (without non-linear activation) to the top layer of the model and finetune it exclusively for particular tasks to evaluate the quality of the embeddings produced by the pre-trained models.
Additionally, we provide the full finetune result of jTrans as it performs best during linear probing.
To showcase the transferability of our model, we do not finetune \sysname. Instead, we utilize the model without any further training for comparison with the baselines.
% Similarly, the zero-shot result of jTrans is also provided for comparsion as it is also reported in the original paper.

% To ensure a comprehensive evaluation, we design six different evaluation tasks: \texttt{O0-O3}, \texttt{O1-O3}, \texttt{O2-O3}, \texttt{O0-Os}, \texttt{O1-Os}, and \texttt{O2-Os}.
The results of our experiments are presented in Table~\ref{tab:binarycorp-10000}. We use Recall@1 and MRR (mean reciprocal rank) as metrics, which are used in previous work~\cite{wangJTransJumpawareTransformer2022}. All baselines are evaluated with a pool size of 10,000, which means there is only one positive sample among 10,000 functions, which are sampled randomly from the whole dataset. Our results indicate that \sysname outperforms the closest baseline competitor by 0.244 for the MRR metric and by over 26.5\% for the recall@1 metric on average across different tasks, which is even a finetuned model.

% \begin{equation}
% \small
%   \begin{aligned}
%   \text{Recall}@k = \frac{1}{|F|} \sum_{f_{i}\in F}\mathbb{I}&(\text{Rank}_{g_{i}} \le k), \text{ where } \mathbb{I}(x)=
%     \begin{cases}
%     0, & x = \text{False}\\
%     1, & x = \text{True}
%     \end{cases}  \\
%   & \text{MRR} = \frac{1}{|F|} \sum_{f_{i} \in F} \frac{1}{Rank_{g_{i}}}
%   \end{aligned}
%   \label{eq:recall-mrr-define}
% \end{equation}

% \begin{equation}
%     \small \text{Recall}@k = \frac{1}{|F|} \sum_{f_{i}\in F}\mathbb{I}(\text{Rank}_{g_{i}} \le k), \text{ where } \mathbb{I}(x)=
%     \begin{cases}
%     0, & x = \text{False}\\
%     1, & x = \text{True}
%     \end{cases}
%     \label{eq:recall}
% \end{equation}

% \begin{equation}
%     \small \text{MRR} = \frac{1}{|F|} \sum_{f_{i} \in F} \frac{1}{Rank_{g_{i}}}
%     \label{eq:mrr}
% \end{equation}

These results demonstrate that even in a zero-shot format, \sysname can significantly outperform state-of-the-art approaches like jTrans and Trex. We also evaluate jTrans in a zero-shot format, where it is pre-trained with the MLM and JTP approach, and the result is shown as jTrans (Zero Shot) in Table~\ref{tab:binarycorp-10000}. The results show that \sysname outperforms jTrans in the zero-shot format by 0.453 for the MRR metric and by over 49.6\% for the recall@1 metric. This reveals that contrastive learning with natural language supervision yields better assembly code representations compared to assembly language models that employ only self-supervised learning.

\iffalse
Furthermore, we explore the influence of different poolsizes while keeping other settings constant. We test poolsizes of
% \texttt{2}, ~\texttt{10}, ~\texttt{32}, ~\texttt{64}, ~\texttt{128}, ~\texttt{256}, ~\texttt{512}, ~\texttt{1,024}, ~\texttt{2,048}, ~\texttt{4,096}, ~\texttt{8,192}, and \texttt{10,000}
$2^i, i \in [1, 13]$ and \texttt{10,000}
, recording the recall@1 metric for each. Figure~\ref{fig:BinaryCorp-Recall@1-Poolsize} presents the results, clearly indicating that as the poolsize increases, the performance of all baselines is inferior to that of \sysname. Moreover, the decline in \sysname's performance is not as significant, suggesting that \sysname is more effective in handling large poolsize settings, which highlights the practical value of our model.

\begin{figure}[t!]
  \centering
  \setlength{\abovecaptionskip}{2mm}
  \scalebox{0.95}{
  \includegraphics[width=1\linewidth]{images/BinaryCorp.pdf}
  }
  \caption{The performance of different binary similarity detection methods on BinaryCorp. The x-axis is logarithmic and denotes the poolsize.}
  
  \label{fig:BinaryCorp-Recall@1-Poolsize}
\end{figure}
\fi

\subsubsection{Crypto Identification}
\label{subsubsec:crypto-identification}
The crypto identification task is used to identify which cryptographic algorithm the assembly code belongs to.
In this experiment, we first filter out the packages that are recognized as encryption algorithms from the Ubuntu repository and extract the functions.
% and then obtain the corresponding binary files compiled with different compilers and optimization options from the dataset engine.
% We use IDA Pro to decompile stripped binary files and extract assembly code representations of functions from them.
For each function, we classify it as an encryption algorithm based on its name and the meta information contained in the corresponding package. It is worth noting that, we ensure the functions are not presented in the training set of constrastive learning to avoid information leakage. We filter out functions with the number of basic blocks smaller than 3, which means that these functions possibly do not contain useful information. Finally, we obtain the dataset containing 17 types of encryption algorithms and about 70K functions.  
% As for the training and testing of linear probe, we split our dataset by packages and use it for training.
% This approach ensures each package in the dataset is independent and not correlated with other packages.
% We believe that this approach is crucial for maintaining the confidentiality of the data and preventing any inadvertent leakage of information.
For comparison, jTrans~\cite{wangJTransJumpawareTransformer2022}, Trex~\cite{peiTrexLearningExecution2021}, and PalmTree~\cite{liPalmTreeLearningAssembly2021} are used as pre-trained baselines.

% \wh{Capitalize the first character for ratio}

\begin{figure}
    \centering
    % \subfloat[\centering Accuracy]{{\includegraphics[width=0.4\linewidth]{images/percent_encryption-identification.pdf} 
    % \label{fig:acc-crypto-percent}
    % }}%
    % \qquad
    % \subfloat[\centering Acc. Distribution]{{\includegraphics[width=0.4\linewidth]{images/encryption_volin.pdf}
    % \label{fig:acc-violin-crypto}
    % }}%
    \setlength{\abovecaptionskip}{2mm}
    \includegraphics[width=.8\linewidth]{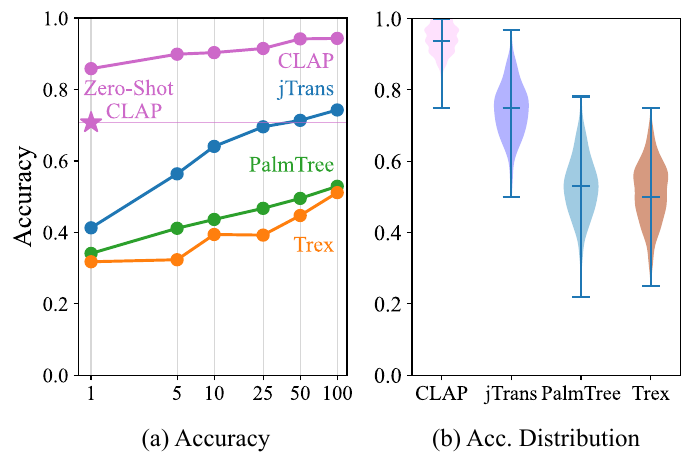}
    \caption{(a) Accuracy of jTrans, PalmTree and \sysname in crypto identification task with different percentages of the training set. The x-axis shows the ratios of the training set (\%). (b) Accuracy distribution with full training set finetuning.}
    \label{fig:acc-crypto-with-violin}%
    \vspace{-2mm}
\end{figure}

To adapt to the crypto identification downstream tasks, we evaluate the models using the linear probe method.
% method by adding a linear classifier on top of the output of model.
We finetune the models with 1\%, 5\%, 10\%, 25\%, 50\%, and 100\% of training data, respectively. Figure~\ref{fig:acc-crypto-with-violin} shows the performance of \sysname and each baseline
% , which are finetuned using different percentages of training data. 
We can see that \sysname achieves a high accuracy of 0.94 on the dataset with 50\% training set and all training data. And even when trained on only 1\% of the data, \sysname's performance surpasses the results of the other two models trained on all training data. This strongly proves our model's excellent generalization ability and effectiveness. We also point out the zero-shot performance (More explanation in Section~\ref{subsec:zero-shot-inference-for-downstream-tasks} and~\ref{subsec:zero-few-shot}) in Figure~\ref{fig:acc-crypto-with-violin}a.

We also record the accuracy of each batch during the evaluation process by the model that trained on all training data and explore the data distribution characteristics of experimental results by drawing violin plots. From Figure~\ref{fig:acc-crypto-with-violin}b, it can be seen that for \sysname, the accuracy is mainly concentrated in a higher range of values. This indicates that our model can accurately classify samples in most cases. 
% For the other baselines, the data distribution shows a symmetrical shape with uniform or normal distribution characteristics, and most of the data values are concentrated around the median. 
Compared to the baselines, \sysname has better stability and more concentrated results.

\subsubsection{Protocol Categorization}
\label{subsubsec:protocol-categorization}

% In this experiment, we use derived functions from protocol libraries to classify protocols and build efficient datasets. We crawl and filter protocol-related development libraries like crypto identification from the Ubuntu repository and extract functions from their export tables. By using the longest match algorithm to match a single protocol tag to function names that may contain protocol names,
Similar to the preprocessing and experiment in Section~\ref{subsubsec:crypto-identification}, we use the packages marked as protocol and obtain a dataset of 736K functions, containing 18 protocol types.
% We also adopt the linear probe method for protocol categorization and finetune with different percentages of training sets, respectively.

\begin{figure}
    \centering
    % \subfloat[\centering Accuracy]{{\includegraphics[width=0.45\linewidth]{images/percent_protocol-identification.pdf} 
    % \label{fig:acc-protocol-percent}
    % }}%
    % \qquad
    % \subfloat[\centering Acc. Distribution]{{\includegraphics[width=0.45\linewidth]{images/protocol_volin.pdf} 
    % \label{fig:acc-violin-protocol}
    % }}%
    \setlength{\abovecaptionskip}{2mm}
    \includegraphics[width=.8\linewidth]{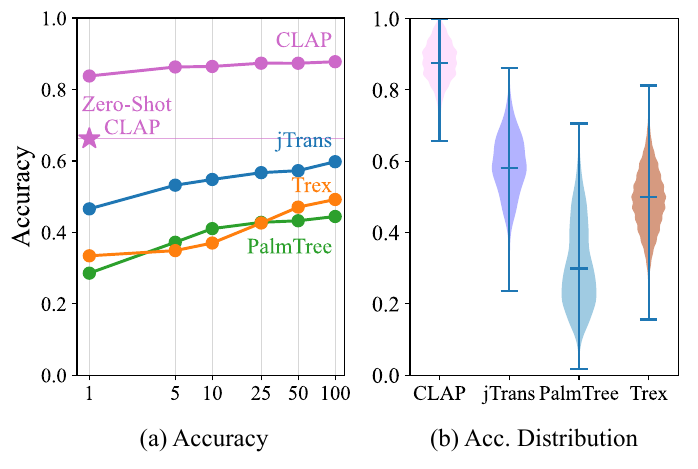}
    \caption{(a) Accuracy of \sysname, jTrans, PalmTree and Trex in protocol categorization task with different percentages of the training set. The x-axis shows the ratios of the training set (\%). (b) Accuracy distribution with full training set finetuning.}
    \label{fig:acc-protocol-with-violin}%
    \vspace{-2mm}
\end{figure}

Figure~\ref{fig:acc-protocol-with-violin}a presents the result of accuracy. It indicates that our method, in the crypto identification task, necessitates a mere 1\% of data for finetuning, enabling it to surpass the performance of other baselines finetuned on the full dataset. Furthermore, even the zero-shot approach alone outperforms or approaches other models finetuned on the full dataset, highlighting our model's superior capability in obtaining an outstanding assembly code representation.
The violin plot of accuracy distribution is shown in Figure~\ref{fig:acc-protocol-with-violin}b, showing an impressive and stable distribution similar to the one in Section~\ref{subsubsec:crypto-identification}.

% We also collect the accuracy of each batch in the evaluation process of the model trained on full training data and visualize it using a violin plot in Figure~\ref{fig:acc-protocol-with-violin}b. The plot shows that the accuracy distribution of \sysname for protocol categorization is more concentrated and stable compared to that of jTrans and PalmTree, whose accuracy distributions are more sparse and less accurate. It indicates that \sysname has learned more stable knowledge and has a better understanding of assembly code.

The conclusion can be drawn by combining the analysis of the aforementioned aspects, indicating that our model acquires a superior representation of assembly code and demonstrates greater stability in its learning results.

\subsection{Transferability Evaluation (RQ2)}\label{subsec:zero-few-shot}

Zero-shot learning allows the model to make predictions or solve novel tasks without having seen any examples from the task's specific class beforehand, while few-shot learning enables the model to adapt quickly to new tasks with only a small number of examples. In this evaluation, we strive to assess the model's zero-shot and few-shot capabilities to validate its transferability performance.

\subsubsection{Zero Shot}

A major benefit of aligning assembly code with natural language explanation is the capacity to seamlessly interact with assembly code using natural language. We explore the model's zero-shot capabilities to highlight this benefit.
% Throughout the zero-shot assessment, natural language processing is employed for communication with ASM. 
% We use specific prompts for the downstream task, including Crypto Identification~\ref{subsubsec:crypto-identification} and Protocol Categorization~\ref{subsubsec:protocol-categorization}.
% \wh{rewrite this part}
% We use the prompts in Table~\ref{lst:prompt-crypto-protocol} as prompt for Crypto Identification and Protocol Categorization.
We use the method introduced in Section~\ref{subsec:zero-shot-inference-for-downstream-tasks} to conduct the zero-shot experiments.
% \elsa{todo: Add prompt in github}
% For each single-classification task, we generate all possible sentences by replacing \texttt{\{class\_name\}} with the labels of downstream tasks and create text embeddings corresponding to their prompts. We find the text explanation with the closest embedding distance in the asm, using the associated label as the few-shot learning outcome.

% \begin{table}[t!]
% \centering
% \fontsize{10.0pt}{\baselineskip}\selectfont
% \linespread{1.0}\selectfont
% \begin{boxedminipage}{1.0\columnwidth}
% % In-context example 1
% \mybox{
% \small
% This function is a function implements encryption \{class\_name\} algorithm.
% Tags: \{class\_name\}, encryption

% This function is a function implements protocol \{class\_name\} algorithm.}
% \end{boxedminipage}
% \caption{Prompt for Crypto Identification (upper) and Protocol Categorization (lower)}
% \label{lst:prompt-crypto-protocol}
%     \vspace{-2mm}
% \end{table}

% \begin{table}[h!]
% \centering
% \fontsize{10.0pt}{\baselineskip}\selectfont
% \linespread{1.0}\selectfont
% \begin{boxedminipage}{1.0\columnwidth}
% % In-context example 1
% \mybox{
% \small
% This function is a function implements protocol \{class\_name\} algorithm.
% }
% \end{boxedminipage}
% \caption{Prompt for Protocol Categorization}
% \label{lst:prompt-protocol}
% \end{table}

\subsubsection{Few Shot}

We acknowledge the challenges in obtaining data for binary tasks, as they often necessitate manual analysis to acquire a limited number of samples. To simulate this, we provide each model with a small set of samples and employ the linear probe technique. To compare the few-shot capability of each baseline, we train the model using the linear probe method with a small dataset, in which each label has 1, 2, 4, 8, or 16 samples. Due to the limited number of samples, the selection of the samples may have a significant impact on the final results, so we repeat the experiments for each sample size five times.
% , present in the Figure~\ref{fig:acc-sample} as grey lines, and average these five outcomes, depict in colorful lines in the figure.

\subsubsection{Results}

\begin{figure}
    \centering
    % \subfloat[\centering Crypto]{{\includegraphics[width=0.45\linewidth]{images/sample_encryption-identification.pdf}
    % \label{fig:acc-crypto-sample}
    % }}%
    % \qquad
    % \subfloat[\centering Protocol]{{\includegraphics[width=0.45\linewidth]{images/sample_protocol-identification.pdf}
    % \label{fig:acc-protocol-sample}
    % }}%
    \setlength{\abovecaptionskip}{2mm}
    \includegraphics[width=0.8\linewidth]{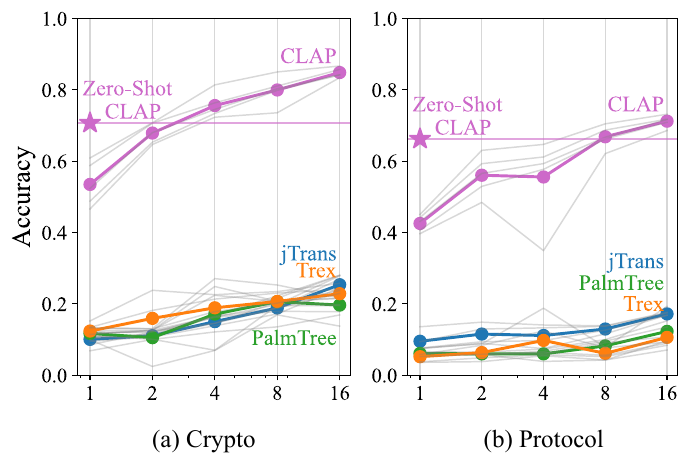}
    \caption{Accuracy of different few samples on Crypto Identification (a) and Protocol Categorization (b). The x-axis shows the number of labeled training examples per class. Each few-show experiment is conducted 5 times. The grey lines show the accuracy of an individual trial. The colored lines show the average accuracy across different trials.}
    \label{fig:acc-sample}%
    \vspace{-2mm}
\end{figure}

The results in the Figure~\ref{fig:acc-sample} indicate exciting results.
%Although CLAP was trained with only one sample per label, the accuracy reached up to 0.38?\wh{need a comfirm}.
% 当每一个label只使用一个sample来进行train的时候，accuracy往往不如直接进行zero-shot，可能是由于overfitting，说明了我们的模型能够在无样本的时候就取得可用的效果，只有在每一个类别的样本数量足够多的时候时，few shot的结果才能超过zero-shot。我们将这个归结为zero-shot和few-shot的区别，在进行zero-shot的时候，自然语言可以直接与对应的汇编代码的特征进行直接的交互，而普通的few-shot的方式只能通过间接的方式来与特征进行交互，使得在样本较少的时候，无法学习得到比较好的汇编代码的特征。
When using only a single sample for training each label, accuracy often falls short compared to zero-shot learning, potentially due to overfitting. This demonstrates that our model can achieve satisfactory results even without samples. It is only when there is an ample number of samples for every category that few-shot learning results surpass those of zero-shot learning.
We attribute this difference to the way each method interacts with assembly code features: while zero-shot learning allows natural language to directly engage with these features, conventional few-shot learning relies on indirect interaction, resulting in difficulty when learning with limited samples.
In contrast, the other baselines are even struggling to learn anything. The accuracy for the baselines is all below 0.3.

Therefore, the above results significantly demonstrate that \sysname has excellent zero-shot and few-shot capability, thus exhibiting remarkable transferability even in scenarios where data is severely constrained or absent.
% The experiments also show that it is easy to transfer \sysname to various downstream tasks and achieve good results.

% \begin{figure}
%     \centering
%     \begin{subfigure}{.5\textwidth}
%       \centering
%       \includegraphics[width=.4\linewidth]{images/sample_encryption-identification.pdf}
%       \caption{Crypto Identification}
%       \label{fig:acc-crypto-sample}
%     \end{subfigure}%
%     \begin{subfigure}{.5\textwidth}
%       \centering
%       \includegraphics[width=.4\linewidth]{images/sample_protocol-identification.pdf}
%       \caption{Protocol Categorization}
%       \label{fig:acc-protocol-sample}
%     \end{subfigure}
    
%     \caption{Accuracy of different few samples on Crypto Identification and Protocol Categorization}
%     \label{fig:enter-label}
% \end{figure}

% \begin{figure}[t]
%   \centering
%   \includegraphics[width=0.9\linewidth]{images/sample_encryption-identification.pdf}
%   \caption{Accuracy of jTrans, PalmTree and \sysname in crypto identification task with few samples.}
%   \label{fig:acc-crypto-sample}
% \end{figure}

% \begin{figure}[t]
%   \centering
%   \includegraphics[width=0.9\linewidth]{images/sample_protocol-identification.pdf}
%   \caption{Accuracy of jTrans, PalmTree and \sysname in protocol categorization task with few samples.}
%   \label{fig:acc-protocol-sample}
% \end{figure}

\subsection{Explaining Natural Language's Role in Assembly Representation Learning (RQ3)}
To investigate the role of natural language supervision in assembly code representation, we conduct some intrinsic evaluations to study how the assembly code representation changes before and after alignment with natural language representations.

\subsubsection{Representation Visualization}
We first utilize t-SNE~\cite{van2008visualizing} to visualize the assembly representation and enhance our understanding of how the \sysname model effectively learns from natural language supervision. As a tool renowned for transforming high-dimensional data into more comprehensible two or three-dimensional formats, t-SNE provides a clear and intuitive view of data relationships, thereby illustrating the nuanced alignment achieved by our model.
% To visually demonstrate the efficiency of the \sysname model in aligning functions, we employ the t-SNE (t-distributed Stochastic Neighbor Embedding) algorithm for visualization purposes~\cite{van2008visualizing}. t-SNE is a widely used technique for data visualization, effectively reducing high-dimensional data to two or three dimensions and facilitating the intuitive observation of similarities and differences.

\begin{figure*}[t!]
    \centering
    \setlength{\abovecaptionskip}{2mm}
    \includegraphics[width=0.95\linewidth]{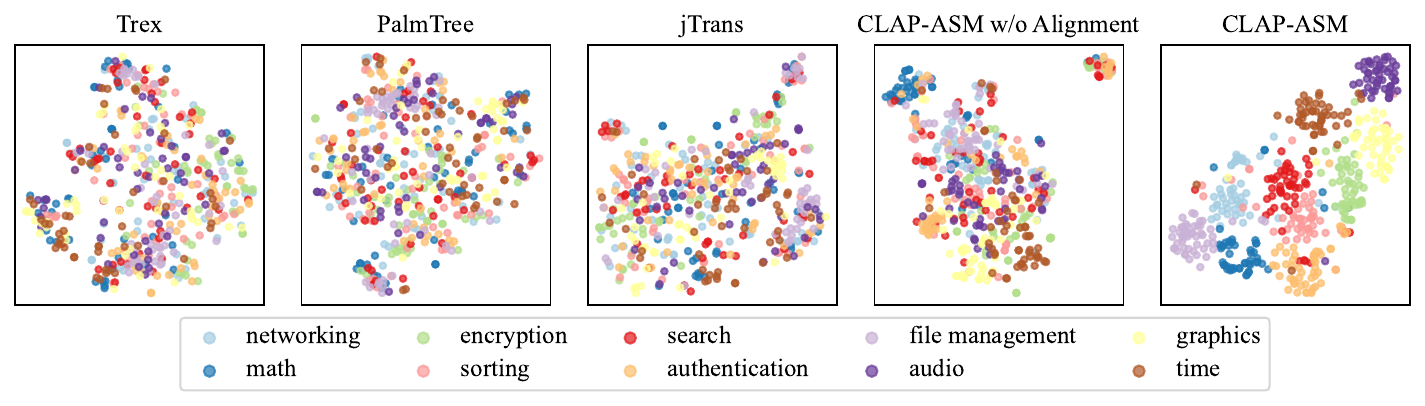}
    \caption{Intrinsic Evaluation: t-SNE visualization of the embedding of assembly code from different models.}
    \label{fig:tsne}
    \vspace{-2mm}
\end{figure*}

For the experiment, we manually select ten functionality labels, including \texttt{time}, \texttt{encryption}, \texttt{authentication}, \texttt{search}, \texttt{math}, file \texttt{management}, \texttt{graphics}, \texttt{networking}, \texttt{sorting}, and \texttt{audio}. From real-world programs, we carefully handpick 50 functions that exemplify each functionality, resulting in a dataset called the "Intrinsic Dataset" comprising 500 functions.
Next, the functions are encoded by five models: Trex, PalmTree, jTrans, and the \asmencoder model both with and without alignment. The t-SNE algorithm then reduces these representations to two dimensions, facilitating visualization shown in Figure~\ref{fig:tsne}.
% Subsequently, the functions are encoded using five models:
% Trex, PalmTree, jTrans, \asmencoder model with or without alignment.
% Following the encoding process, the t-SNE algorithm is utilized to reduce the dimensionality of the resulting representations to two dimensions. This enables the visualization of the outcomes, as depicted in Figure~\ref{fig:tsne}.

In the visualization, we observe that the embeddings generated by the \asmencoder without alignment, as well as those generated by Trex, PalmTree, and jTrans all display a nearly disordered distribution. This pattern of disarray is consistent across these models, indicating a lack of high-level program semantics in their embeddings. 
% In contrast, the embeddings produced by the \textencoder, due to the intrinsic nature of natural language explanations, exhibit clear differentiation into multiple clusters after dimensionality reduction. This result emphasizes the text model's robust ability to discern functions based on their natural language explanations, thereby validating the promise of alignment.
In contrast, after natural language supervision which aligns with text embedding, the \asmencoder model's output can be reduced to the same vector space according to the functionality, displaying similar distribution characteristics. This visual evidence underscores the alignment effect between our model and the natural language model, implying the potential for manipulating assembly language models using natural language directly in a zero-shot scenario.

\subsubsection{Insights of Natural Language Supervision}
After the implementation of contrastive learning training enhanced by natural language supervision, we observe a notable performance improvement. This enhancement can be attributed to the introduction of essentially unlimited labels via sophisticatedly trained text encoders. Unlike contrastive learning on the solitary modality of assembly code, the incorporation of natural language signals offers pivotal anchors throughout the training process. This approach not only enriches the training context but also cultivates features with superior transferability.

To elucidate our findings, consider a scenario with two cryptographic algorithms' assembly code implementations: RSA, an asymmetric encryption algorithm, and MD5, a hash function. Contrastive learning, solely reliant on assembly code, typically enables the model to identify similarities within RSA and MD5 implementations independently, but it struggles to link the two. 

Introducing natural language as a supervisory signal changes this dynamic. For example, the natural language descriptions ``RSA, a modern asymmetric encryption algorithm in cryptography'' and ``MD5, a widely used cryptographic hash function'' serve as multifaceted labels. These labels associate RSA with the terms `cryptography, asymmetric encryption, RSA' and MD5 with `cryptography, hash algorithm, MD5', thereby creating a connective thread between the two under the broad theme of cryptography. This highlights the pivotal role of natural language in establishing meaningful connections in assembly code representations.

\subsection{Dataset Engine Evaluation (RQ4)}
\label{subsec:data-engine-evaluation}

In this section, we evaluate the quality of the explanation generated by GPT and Shadow Mode, i.e., whether and how well the explanation matches the source code.

\subsubsection{Quality of Source Explanation}

To assess the quality of source code explanations, we engage five domain experts to independently review and score the explanations from GPT-3.5 and the shadow model. The evaluation used a clear scoring system with four different levels of explanatory quality, ranging from `exemplary' through `acceptable' with minor omissions, to `substandard' with most key points missed, and culminating in `poor' with essentially no accurate explanation of code functionality. The experts will assign the score from 4 (best) to 1 (worst) respectively.

% \elsa{todo: clarify about the error rate}

\begin{figure}
    \centering
    % \subfloat[\centering Human Score Distribution]{{\includegraphics[width=0.38\linewidth]{images/gpt-shadow-model-score.pdf} 
    % \label{fig:human-score}
    % }}%
    % \qquad
    % \subfloat[\centering Human Evaluation]{{\includegraphics[width=0.52\linewidth]{images/human-eval.pdf}
    % \label{fig:human-eval}
    % }}%
    \setlength{\abovecaptionskip}{2mm}
    \includegraphics[width=1\linewidth]{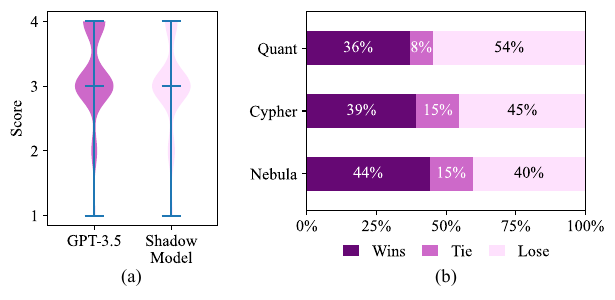}
    \caption{
(a) Human evaluation of the explanations of each model by scoring 4 (best) to 1 (worst) individually.\\
(b) Human evaluation by comparing the explanations from each model side by side, only codename is provided.
Win signifies that the user ranks the model above GPT-3.5, while lose implies a lower ranking. A tie means that users perceive both models to have comparable performance.
Quant: LLaMA 13B model trained on a 50K dataset. Cypher: LLaMA 13B model trained on a 1.3M dataset. Nebula: LLaMA 30B model trained on a 50K dataset.}
    \label{fig:huamn-score-and-human-eval}%
    \vspace{-4mm}
\end{figure}

We selected 50 explanations from each model at random for evaluation by human experts, without disclosing the source of each explanation. The score distribution, as depicted in Figure~\ref{fig:huamn-score-and-human-eval}a, primarily shows both models frequently achieving a score of 3, indicating an acceptable level. Notably, both models exhibit a relatively low rate of critical errors, reflected in the 1-point scores. However, GPT-3.5 slightly outperforms with a marginally higher number of 4-point scores, as evidenced by the average scores. GPT-3.5 achieves an average score of 3.14, modestly surpassing the shadow model's average of 3.00. While these results are not outstanding, they are deemed acceptable, suggesting the viability of using language-based supervision as an effective signal for contrastive learning.
% We randomly pick 50 explanations for the human experts to review from each model without telling them which model produces which explanation.
% The score distribution is shown in Figure~\ref{fig:huamn-score-and-human-eval}a. The distribution indicates that in the majority of cases, both models achieved a score of 3, i.e. `substandard'. And they both have a relatively low error rate, i.e. 1-point score. However, GPT-3.5 received a marginally higher number of 4-point  scores, which is also reflected in the average values.
% On average, the GPT-3.5 model scores 3.14, slightly better than the shadow model, which scores 3.00. The results while not exemplary, but still acceptable, thus signaling potential efficacy as a language-based supervision signal for constrastive learning.

Further quantitative analysis via T-Test comparing mean scores of the two models yields a p-value of 0.0036, which indicates that explanations generated by GPT-3.5 are of significantly higher quality than those produced by the shadow model. Noteworthy is that despite the shadow model not achieving parity with the GPT-3.5, does not significantly trail in the Section~\ref{sec:quality-of-shadow-model} where the experts can compare the explanations from different models side by side.

\subsubsection{Quality of Shadow Model}
\label{sec:quality-of-shadow-model}

To compare the quality of source code explanation from the shadow model and GPT-3.5, we employ existing LLM assessment methods~\cite{zhouLIMALessMore2023}. In particular, We design an online website, in which we provide different explanations from various models for the same source code, and let users rank them. This approach enables us to gauge the wins, losses, and ties among the models in comparison to GPT-3.5.
During the human evaluation, we aim to prevent model information leakage (specifically, model size and training set size) while still obtaining user feedback (allowing users to assess the output quality of various models).

The final results shown in Figure~\ref{fig:huamn-score-and-human-eval}b reveal some discrepancies between GPT-3.5 and the other models in terms of human perception; however, GPT-3.5 does not exhibit significant superiority. Each model experiences both wins and losses. Notably, the 30B model, with the highest parameter count, outperforms the two 13B models, aligning with expectations for larger models. Furthermore, the model trained on the 1.3M dataset encounters a wider array of functions, allowing it to tackle more complex tasks, while the model from the 50K training set fares the worst in human evaluations. Nevertheless, it is crucial to stress that the latter model still competes with GPT-3.5 in terms of code comprehension and interpretation. Due to increased GPU memory requirements and a reduced batch size associated with the 30B model's hosting capability, along with slower inference speeds from larger models, we opt for the 13B LLaMA model from the 1.3M dataset for extensive inference to strike a balance between performance and inference overhead. % Consequently, we get the final text-side data augmentation used for contrastive learning.

\section{Broader Impacts}
% In the experiment, we conduct numerous evaluations to validate the performance of \sysname on various downstream tasks. 

The experimental results showcased above highlight the notable capabilities of \sysname, yet its potential extends further. By bridging the gap between natural language and assembly code, \sysname facilitates support for zero-shot learning through the use of open vocabulary prompts. This feature enhances \sysname's applicability in more scenarios, offering a wider scope of utility.

% Considering the challenges of constructing standardized test sets for diverse tasks, we have decided to present our \sysname's capabilities through case studies. Through these case studies, we delve into and showcase the application and effectiveness of our model in real-world scenarios. Specifically, we have selected challenging real-world problems as cases to demonstrate the advantages and potential of our model in addressing them.
In many potential real-world semantic analysis tasks within the realm of binary analysis, the construction of datasets is exceptionally challenging, which has significantly impeded the application of assembly code representation learning. Due to these difficulties in dataset creation, this section analyzes the broader impact of \sysname, through various case studies. These studies not only demonstrate the practical effectiveness of \sysname but also highlight its advantages in zero-shot learning scenarios. This approach showcases the powerful potential of our model in handling complex scenarios where traditional data-driven methods may fall short.

\subsection{Fine-grained Assembly Semantics}

\label{subsec:sort-algorithm-classification}
\begin{figure}[t]
    \centering
    \setlength{\abovecaptionskip}{2mm}
    \includegraphics[width=0.9\linewidth]{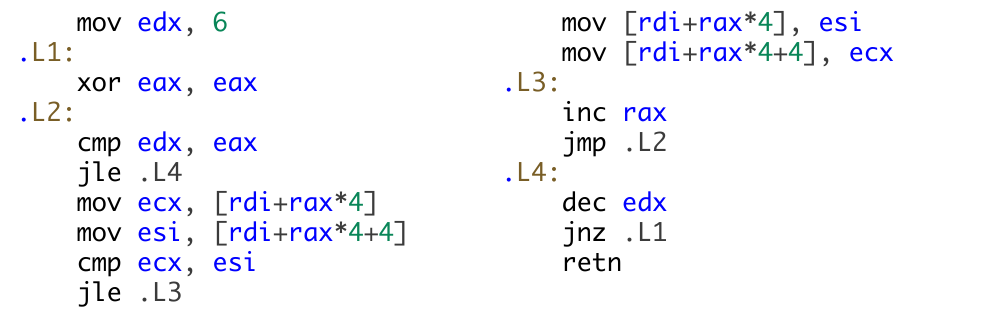}
    \caption{Assembly code of a bubble sort algorithm}
    \label{fig:bubblesort}
    \vspace{-2mm}
\end{figure}

Our first case study focuses on examining the model's potential comprehension of fine-grained program semantics. We choose sorting algorithms, taking the bubble sort algorithm as an example. We investigate whether the model can recognize the given assembly code as the bubble sort algorithm shown in Figure~\ref{fig:bubblesort}. Our input prompts include ten labels: bubble sort, selection sort, insertion sort, merge sort, quick sort, heap sort, radix sort, bucket sort, shell sort, and counting sort, using prompts like ``This is a \texttt{XX} sorting algorithm'' (replacing \texttt{XX} with various sorting algorithms). Employing a zero-shot inference approach, we search for the prompt that best matches the provided assembly code. Experimental results demonstrate that our model can correctly identify the bubble sort algorithm from the ten options, showcasing its ability to comprehend fine-grained program semantics.
% Our first case study focuses on examining the model's comprehension of fine-grained program semantics. We choose sorting algorithms, taking the bubble sort algorithm as an example. We investigate whether the model can recognize the given assembly code as the bubble sort algorithm shown in Listing~\ref{fig:bubblesort}. Our input prompts include ten labels, including Bubble Sort, Selection Sort, Insertion Sort, Merge Sort, Quick Sort, Heap Sort, Radix Sort, Bucket Sort, Shell Sort, and Counting Sort, and utilize the prompt like "This is a bubble sort algorithm." By employing the inference approach, we search for the most appropriate prompt that aligns with the provided assembly code. Experimental results demonstrate that our model can correctly select the bubble sort algorithm from the ten options, with a softmax prediction result of 0.179, surpassing the second-best result of 0.128, thereby showcasing the model's ability to comprehend sorting algorithms.

% \begin{center}
% \begin{minipage}{0.8\linewidth}
% \begin{lstlisting}[label=lst:bubblesort,basicstyle=\scriptsize\ttfamily\bfseries,caption=Assembly code of a bubble sort algorithm, language={[x86masm]Assembler},escapechar=!, frame=single]
%     mov    edx, 6
% .L1:
%     xor    eax, eax
% .L2:
%     cmp    edx, eax
%     jle    .L4
%     mov    ecx, [rdi+rax*4]
%     mov    esi, [rdi+rax*4+4]
%     cmp    ecx, esi
%     jle    .L3
%     mov    [rdi+rax*4], esi
%     mov    [rdi+rax*4+4], ecx
% .L3:
%     inc    rax
%     jmp    .L2
% .L4:
%     dec    edx
%     jnz    .L1
%     retn
% \end{lstlisting}
% \end{minipage}
% \end{center}

\subsection{Malware Behaviour Classification}

% The case study on bubble sort showcases the model's ability in algorithm analysis. However, its application scope is limited. 
In the second case study, we focus on malicious code analysis to demonstrate the model's capabilities in real-world tasks. We manually reverse-engineer functions from malicious samples and find a malicious screenshot function\footnote{Function \texttt{sub\_10001170} in the sample, which can be found at \href{https://www.virustotal.com/gui/file/cd677242197cdc89d7b8e2e3056030fe2bb9b384c95a7a027a7eee8182b8426f/detection}{VirusTotal (url here)}}. We construct prompts encompassing common functionalities that may be used in malware, including a screenshot, auto-start, backdoor, download, upload, rootkit, anti-detect, anti-debug, password brute force, and file hijack. We use \sysname to zero-shot differentiate the malicious assembly function from the above categories. The experiment result shows that our model accurately recognizes the function as being related to screenshots, highlighting the practical capability of our model in real-world malicious code analysis.

% Through these case studies, we have demonstrated the broad applicability of our model in more potential scenarios and its ability to perform different tasks without any further training by natural language prompts. This reinforces the efficacy and promise of \sysname.

\iffalse
\subsection{Assembly Code Search}
\elsa{too short} We are considering another potential application, which is to assist in reverse engineering by using natural language to search for relevant assembly code. Reverse engineers often spend a lot of time analyzing the framework and business logic of a program to find its key logic. Using natural language to search for relevant assembly code can greatly improve productivity.
\fi

\section{Limitations}
While our study presents significant advancements with \sysname, we recognize certain limitations and propose directions for future research.
First, our current approach does not fully harness the wealth of information available in source code.
Future studies could explore methodologies that synergize natural language with source code. This could involve developing more sophisticated algorithms that deeply integrate source code semantics and structures, potentially leading to even more robust models.

Second, our dataset engine, which relies on compiling packages from Ubuntu and using GPT-3.5 for source code explanations, may introduce biases that could potentially degrade performance. To address this, future research could diversify the data sources, including different compilation environments or platforms and utilizing varied language models for source code explanations.

Lastly, the current scope of semantic analysis tasks discussed is limited. There is substantial scope for exploring a broader array of application scenarios, such as assembly code search with natural language. This could include delving into more complex and varied semantic analysis tasks, thereby expanding the model's utility and uncovering new avenues in binary code analysis.

\section{Conclusion}

In conclusion, we have presented \sysname, a novel method for learning assembly code representations through natural language supervision. Our approach successfully bridges the gap between assembly code and natural language representations. The \sysname model achieves remarkable transferability in binary analysis and outperforms state-of-the-art solutions in various tasks.
This study highlights the potential of learning assembly code representation with natural language supervision, unlocking new possibilities for assembly code analysis and representation learning. We believe that our method opens up new avenues for research and offers a promising starting point for future work.

% \section{Data Availability}

% The \sysname model, two-stage pre-training code, prompt used during dataset construction, and case studies are available at \url{https://anonymous.4open.science/r/CLAP-7283}.

% \newpage

% %-------------------------------------------------------------------------------
% \section*{Acknowledgments}
% %-------------------------------------------------------------------------------

% The USENIX latex style is old and very tired, which is why
% there's no \textbackslash{}acks command for you to use when
% acknowledging. Sorry.

% %-------------------------------------------------------------------------------
% \section*{Availability}
% %-------------------------------------------------------------------------------

% USENIX program committees give extra points to submissions that are
% backed by artifacts that are publicly available. If you made your code
% or data available, it's worth mentioning this fact in a dedicated
% section.
% QUERYX: Symbolic Query on Decompiled Code for Finding Bugs in COTS Binaries
% %-------------------------------------------------------------------------------
\bibliographystyle{plain}
\bibliography{main,elsa}

\newpage

% \section*{Appendix}

\iffalse
\begin{center}
\begin{minipage}{0.95\linewidth}
\begin{lstlisting}[label=lst:motivation1asm,basicstyle=\scriptsize\ttfamily\bfseries,caption=Assembly Code of Listing\ref{lst:motivation1}, language={[x86masm]Assembler},escapechar=!, frame=single]
        mov     rax, rsi
        mov     esi, edi
        mov     ecx, edi
        and     esi, 61440
        cmp     esi, 16384
        jne     .L2
        mov     BYTE PTR [rax], 100
        jmp     .L3
.L2:
        cmp     esi, 40960
        jne     .L4
        mov     BYTE PTR [rax], 108
.L3:
        mov     edx, 1
        jmp     .L5
.L4:
        cmp     esi, 8192
        jne     .L6
        mov     BYTE PTR [rax], 99
        jmp     .L3

// ... more assembly here ...

        sbb     ecx, ecx
        and     ecx, -32
        add     ecx, 116
        jmp     .L21
.L20:
        cmp     esi, 1
        sbb     ecx, ecx
        and     ecx, -75
        add     ecx, 120
.L21:
        add     edx, 9
        movzx   esi, di
        movzx   edx, dx
        mov     BYTE PTR [rax+rsi], cl
        mov     BYTE PTR [rax+rdx], 0
        ret
\end{lstlisting}
\end{minipage}
\end{center}
\fi

%%%%%%%%%%%%%%%%%%%%%%%%%%%%%%%%%%%%%%%%%%%%%%%%%%%%%%%%%%%%%%%%%%%%%%%%%%%%%%%%
\end{document}

%% file: config.tex
\usepackage{verbatim}
\usepackage{amsmath}

\usepackage{amssymb}
\usepackage{algorithmic}
\usepackage{algorithm}
\usepackage{makecell}
\usepackage{boxedminipage}

\definecolor{shadecolor}{RGB}{246,246,246}
\newcommand{\mybox}[1]{\vspace{0.3mm}\par\noindent\colorbox{shadecolor}
{\parbox{\dimexpr\textwidth-2\fboxsep\relax}{\vspace{-0.2mm}#1\vspace{-0.2mm}}}\vspace{0.3mm}}

\usepackage{graphicx}
\usepackage{epstopdf}
\usepackage{array}
\usepackage{booktabs}
\usepackage{multirow}
\usepackage{caption} %[justification=centering]
\usepackage{multicol}
\usepackage{tabularx}
\usepackage{framed}
\usepackage{enumitem}
\usepackage{threeparttable}
\usepackage{supertabular}

\graphicspath{{figs/}}
\usepackage{xcolor}

\usepackage{colortbl} % allows rows and columns to be coloured, and even individual cells
\definecolor{jade}{rgb}{0.0, 0.66, 0.42}
\definecolor{carolinablue}{rgb}{0.6, 0.73, 0.89}
\definecolor{dkgreen}{rgb}{0,0.6,0}
\definecolor{dkblue}{rgb}{0,0.4,0.5}
\definecolor{gray}{rgb}{0.5,0.5,0.5}
\definecolor{mauve}{rgb}{0.58,0,0.82}

\definecolor{codegreen}{rgb}{0,0.6,0}
\definecolor{codegray}{rgb}{0.5,0.5,0.5}
\definecolor{codepurple}{rgb}{0.58,0,0.82}
\definecolor{codeblue}{rgb}{0,0,205}
\definecolor{backcolour}{rgb}{245,245,245}

\usepackage{listings}
\usepackage{enumitem}

\lstdefinelanguage
   [x64]{Assembler}     % add a "x64" dialect of Assembler
   [x86masm]{Assembler} % based on the "x86masm" dialect
   {morekeywords={xend, CDQE,CQO,CMPSQ,CMPXCHG16B,JRCXZ,LODSQ,MOVSXD, %
                  POPFQ,PUSHFQ,SCASQ,STOSQ,IRETQ,RDTSCP,SWAPGS, %
                  rax,rdx,rcx,rbx,rsi,rdi,rsp,rbp, %
                  r8,r8d,r8w,r8b,r9,r9d,r9w,r9b, %
                  r10,r10d,r10w,r10b,r11,r11d,r11w,r11b, %
                  r12,r12d,r12w,r12b,r13,r13d,r13w,r13b, %
                  r14,r14d,r14w,r14b,r15,r15d,r15w,r15b}} % etc.

\lstdefinestyle{mystyle}{
    backgroundcolor=\color{backcolour},   
    commentstyle=\color{codegreen},
    keywordstyle=\color{codeblue},
    numberstyle=\tiny\color{codegray},
    stringstyle=\color{codeblue},
    basicstyle=\ttfamily\footnotesize,
    breakatwhitespace=false,         
    breaklines=true,                 
    captionpos=b,                    
    keepspaces=true,                 
    numbers=left,                    
    numbersep=5pt,                  
    showspaces=false,                
    showstringspaces=false,
    showtabs=false,                  
    tabsize=2
}
\usepackage{url}
\usepackage{flushend}
\usepackage{balance}

\usepackage[misc]{ifsym}
\usepackage{bbding}

\usepackage{xspace}

\newcommand{\authnote}[2]{{\bf \textcolor{blue}{#1}: \em \textcolor{red}{#2}}}

\newcommand{\sysname}{{\tt CLAP}\xspace}
\newcommand{\asmencoder}{{\tt CLAP-ASM}\xspace}

\newcommand{\binarycorp}{{\tt BinaryCorp}\xspace}

\newcommand{\elsa}[1]{\authnote{elsa}{#1}}

\usepackage{xspace}

\usepackage{subfig}

 \usepackage[markup=underlined]{changes}